\documentclass[11pt]{article}
\usepackage{graphicx}

\newcommand{\BABARPubYear}    {05}

\newcommand{\BABARConfNumber} {018}
\newcommand{\SLACPubNumber} {11377}

\newcommand{\de}{\ensuremath{\mbox{$\Delta E$}}\xspace}
\newcommand{\fis}{\ensuremath{\mbox{$\mathcal{F}$}}\xspace}
\def\Dztilde   {\ensuremath {\tilde{D}^0}\xspace}

\def\Dzb   {\ensuremath {{\Db}^0}\xspace}
\def\Dztilde   {\ensuremath {\tilde{D}^0}\xspace}

\def\dodstartilde {\ensuremath {\tilde{D}^{(\ast)0}}\xspace}

\def \rb {\ensuremath {r_B}\xspace}

\def \rbs {\ensuremath {r^\ast_B}\xspace}
\def \rbbs {\ensuremath {r^{(\ast)}_B}\xspace}
\def \deltab {\ensuremath {\delta_B}\xspace}

\def \deltabbs {\ensuremath {\delta^{(\ast)}_B}\xspace}
\def \rs {\ensuremath {r_s}\xspace}
\def \rstwo {\ensuremath {r_s^2}\xspace}
\def \deltas {\ensuremath {\delta_s}\xspace}

\def \xbmp {\ensuremath {x_\mp}\xspace}

\def \ybmp {\ensuremath {y_\mp}\xspace}

\def \xbbspm {\ensuremath {x_\pm^{(\ast)}}\xspace}
\def \ybbspm {\ensuremath {y_\pm^{(\ast)}}\xspace}

\def \xsp {\ensuremath {x_{s+}}\xspace}
\def \xsm {\ensuremath {x_{s-}}\xspace}
\def \xspm {\ensuremath {x_{s\pm}}\xspace}
\def \xsmp {\ensuremath {x_{s\mp}}\xspace}
\def \ysp {\ensuremath {y_{s+}}\xspace}
\def \ysm {\ensuremath {y_{s-}}\xspace}
\def \yspm {\ensuremath {y_{s\pm}}\xspace}
\def \ysmp {\ensuremath {y_{s\mp}}\xspace}
\def \xs {\ensuremath {x_{s}}\xspace}
\def \ys {\ensuremath {y_{s}}\xspace}

\def \zpm {\ensuremath {{\bf z}_\pm}\xspace}
\def \zplus {\ensuremath {{\bf z}_+}\xspace}
\def \zminus {\ensuremath {{\bf z}_-}\xspace}
\def \zpdata {\ensuremath {{\bf z}_+^{\rm data}}\xspace}
\def \zmdata {\ensuremath {{\bf z}_-^{\rm data}}\xspace}
\def \p {\ensuremath {{\bf p}}\xspace}

\newcommand{\re}{\ensuremath{\mathop{\rm Re}}}
\newcommand{\im}{\ensuremath{\mathop{\rm Im}}}

\def\bea{\begin{eqnarray}}
\def\eea{\end{eqnarray}}
\def\nn{\nonumber}

\newcommand{\reslinepolvalppp}[4]{\left(#1\pm#2\pm#3\pm#4\right)^\circ}
\newcommand{\gevtwo}{\ensuremath{{\mathrm{\,Ge\kern -0.1em V^2\!/}c^4}}\xspace}

\input pubboard/babarsym

\long\def\inst#1{\par\nobreak\kern 4pt\nobreak
    {\it #1}\par\vskip 10pt plus 3pt minus 3pt}

\begin{document}
{\pagestyle{empty}

\begin{flushright}
\babar-CONF-\BABARPubYear/\BABARConfNumber \\
SLAC-PUB-\SLACPubNumber \\
July 2005 \\
\end{flushright}

\par\vskip 5cm

\begin{center}
\Large \bf Measurement of $\gamma$ in $\Bmp \to D^{(*)} \Kmp$ and $\Bmp \to D \Kstarmp$ decays with a Dalitz analysis of $D \to \KS\pim\pip$
\end{center}
\bigskip

\begin{center}
\large The \babar\ Collaboration\\
\mbox{ }\\
\today
\end{center}
\bigskip \bigskip

\begin{center}
\large \bf Abstract
\end{center}
We present a measurement of the Cabibbo-Kobayashi-Maskawa \CP-violating phase $\gamma$ with a Dalitz plot analysis
of neutral $D$-meson decays to the $\KS\pim\pip$ final state from 
$\Bmp \to D^{(*)} \Kmp$ and $\Bmp \to D \Kstarmp$ decays, 
using a sample of 227 million \BB pairs collected by the \babar\ detector. 
We measure $\gamma = \reslinepolvalppp{67}{28}{13}{11}$, where
the first error is statistical, the second is the experimental systematic uncertainty
and the third reflects the Dalitz model uncertainty. This result suffers from a two-fold ambiguity.
The contribution to the Dalitz model uncertainty due to the description of the $\pi\pi$ S-wave in $\Dz \to \KS \pim \pip$,
evaluated using a K-matrix formalism, is found to be $3^\circ$.

\vfill
\begin{center}
Submitted at the 
International Europhysics Conference On High-Energy Physics (HEP 2005),
7/21---7/27/2005, Lisbon, Portugal
\end{center}

\vspace{1.0cm}
\begin{center}
{\em Stanford Linear Accelerator Center, Stanford University, 
Stanford, CA 94309} \\ \vspace{0.1cm}\hrule\vspace{0.1cm}
Work supported in part by Department of Energy contract DE-AC03-76SF00515.
\end{center}

\newpage
} 

\begin{center}
\small

The \babar\ Collaboration,
\bigskip

B.~Aubert,
R.~Barate,
D.~Boutigny,
F.~Couderc,
Y.~Karyotakis,
J.~P.~Lees,
V.~Poireau,
V.~Tisserand,
A.~Zghiche
\inst{Laboratoire de Physique des Particules, F-74941 Annecy-le-Vieux, France }
E.~Grauges
\inst{IFAE, Universitat Autonoma de Barcelona, E-08193 Bellaterra, Barcelona, Spain }
% ...

\end{center}\newpage

\section{INTRODUCTION}
\label{sec:Introduction}
\CP violation in the Standard Model is described by a single phase
in the Cabibbo-Kobayashi-Maskawa (CKM) quark-mixing matrix~\cite{ref:ckm}.
Although \CP violation in the $B$ system is now well established, further
measurements of \CP violation are needed to overconstrain the 
Unitarity Triangle~\cite{ref:pdg2004} and confirm the CKM model 
or observe deviations from its predictions.
The angle $\gamma$ of the Unitarity Triangle is defined as 
$\gamma\equiv\arg{\left[-V_{ud}^{}V_{ub}^{*}/V_{cd}^{}V_{cb}^{*}\,\right]}$.
Various methods~\cite{ref:GLWADS,ref:DKDalitz} have been proposed to 
extract $\gamma$ using 
$\Bm \to \Dztilde \Km$~\footnote{Reference to the charge-conjugate state is implied here and throughout
the text unless otherwise specified.} decays, all exploiting the
interference between the color allowed $\Bm \to \Dz \Km$ ($\propto
V_{cb}$) and the color suppressed $\Bm \to \Dzb \Km$ ($\propto
V_{ub}$) transitions, when the \Dz and \Dzb are reconstructed
in a common final state. The symbol \Dztilde indicates either a \Dz or a \Dzb meson.
The extraction of $\gamma$ with these decays
is theoretically clean because the main contributions to the
amplitudes come from tree-level transitions.

Among the \Dztilde decay modes studied so far the $\KS \pim \pip$ channel is
the one with the highest sensitivity to $\gamma$ because of the best
overall combination of branching ratio magnitude, $\Dz-\Dzb$
interference and background level. Both \babar~\cite{ref:babar_dalitz} and 
Belle~\cite{ref:belle_dalitz} have reported on a measurement of $\gamma$ based
on $\Bm \to \dodstartilde \Km$ decays with a Dalitz analysis of $\Dztilde \to \KS \pim \pip$. 
Here, the symbol ``$(*)$'' refers to either a $D$ or \Dstar meson.
Belle has recently shown a preliminary result using $\Bm \to \Dztilde \Kstarm$, $\Kstarm \to \KS\pim$~\cite{ref:belle_btodkstar}.
In this paper we report on the update of the $\gamma$ measurement with the
addition of the $\Bm \to \Dztilde \Kstarm$, $\Kstarm \to \KS\pim$ decay mode to the previously used
$\Bm \to \dodstartilde \Km$ channels.

Assuming no \CP asymmetry in $D$ decays and neglecting the 
$\Bmp\to \Dztilde (\KS\pi^\mp)_{\textrm{non}-\Kstar}$ contribution, the $\Bmp \to \Dztilde \Kstarmp$, $\Dztilde \to \KS \pim \pip$,
$\Kstarmp \to \KS \pimp$ decay chain rate $\Gamma_\mp(m^2_-,m^2_+)$ can be \mbox{written as}
\bea
\Gamma_\mp(m^2_-,m^2_+) \propto |{\cal A}_{D\mp}|^2 + \rb^2 |{\cal A}_{D\pm}|^2 + 
   2 \left\{ \xbmp \re[{\cal A}_{D\mp} {\cal A}_{D\pm}^*]+\ybmp\im[ {\cal A}_{D\mp} {\cal A}^*_{D\pm}] \right\} ~,
\label{eq:ampgen1}
\eea
where $m^2_-$ and $m^2_+$ are the squared invariant masses of the $\KS\pim$ and $\KS\pip$ combinations
respectively from the \Dztilde decay, and ${\cal A}_{D\mp} \equiv {\cal A}_{D}(m^2_\mp,m^2_\pm)$, with
${\cal A}_{D-}$ (${\cal A}_{D+}$) the amplitude of the $\Dz \to \KS\pim\pip$ ($\Dzb \to \KS\pip\pim$) decay.
In Eq.~(\ref{eq:ampgen1}) we have introduced the {\it Cartesian coordinates} $\xbmp = \rb \cos(\deltab \mp\gamma)$ and 
$\ybmp=\rb \sin(\deltab \mp\gamma)$~\cite{ref:babar_dalitz}, for which the constraint $\rb^2=\xbmp^2 +\ybmp^2$ holds. Here,
\rb is the magnitude of the ratio of the amplitudes ${\cal A}(\Bm \to \Dzb \Kstarm)$ and ${\cal A}(\Bm \to \Dz \Kstarm)$ 
and \deltab is their relative strong phase.

In the case where a $\Bmp\to \Dztilde (\KS\pimp)_{\textrm{non}-\Kstar}$ component interferes
with $\Bmp\to \Dztilde \Kstarmp$, we write a general parameterization of the decay rates following the
approach proposed in Ref.~\cite{ref:gronau2002},
\bea
\Gamma_\mp(m^2_-,m^2_+) \propto |{\cal A}_{D\mp}|^2+\rs^2|{\cal A}_{D\pm}|^2+ 2 \left\{ \xsmp \re[{\cal A}_{D\mp} {\cal A}^*_{D_\pm}] + 
       \ysmp \im[{\cal A}_{D\mp} {\cal A}^*_{D\pm}] \right\}~,\label{eq:ampgen2}
\eea
where $\xsmp = \kappa \rs \cos(\deltas\mp\gamma)$, $\ysmp =\kappa \rs \sin(\deltas\mp\gamma)$ and $\xsmp^2+\ysmp^2=\kappa^2 \rs^2$, 
with $0\leq \kappa\leq 1$. 
In the limit of a null $\Bmp\to \Dztilde (\KS\pimp)_{\textrm{non}-\Kstar}$ contribution, $\kappa\to 1$, $\rs\to \rb$ and $\deltas\to \deltab$.
The parameterization given by Eq.~(\ref{eq:ampgen2}) is also valid in the case when \rb and \deltab happen to vary within the \Kstar mass window,
and accounts for efficiency variations as a function of the kinematics of the \B decay.

Once the decay amplitude ${\cal A}_{D}$ is known, the Dalitz plot distributions for \Dztilde from \Bm and \Bp decays 
can be simultaneously fitted to $\Gamma_-(m^2_-,m^2_+)$ and $\Gamma_+(m^2_-,m^2_+)$ as given by Eq.~(\ref{eq:ampgen2}), 
respectively. A maximum likelihood technique can be used to estimate the \CP-violating parameters
\xsmp, \ysmp, and \rstwo. Since the parameter \rstwo is also floated our $(\xsmp,\ysmp)$ results do not depend
on any assumption on the amount and nature of the $\Bm\to \Dztilde (\KS\pim)_{\textrm{non}-\Kstar}$ component, 
while on average the statistical uncertainties do not increase. 
Moreover, this general treatment allows us to consider the $\Bm\to \Dztilde (\KS\pim)_{\textrm{non}-\Kstar}$ events like 
$\Bm\to \Dztilde \Kstarm$ signal.

Since the measurement of $\gamma$ arises from the interference term in Eq.~(\ref{eq:ampgen2}), the uncertainty in the 
knowledge of the complex form of ${\cal A}_{D}$ can lead to a systematic uncertainty. Two different models 
describing the $\Dz \to \KS \pim \pip$ decay have been used in this analysis. The first model (also referred to 
as Breit-Wigner model)~\cite{ref:cleomodel}
is the same as used for our previously reported measurement of $\gamma$ on 
$\Bm \to \dodstartilde \Km$, $\Dztilde \to \KS \pim \pip$ decays~\cite{ref:babar_dalitz}, and expresses ${\cal A}_{D}$ as a sum of 
two-body decay-matrix elements and a non-resonant contribution. In the second model (hereafter referred to as the 
$\pi\pi$ S-wave K-matrix model) the treatment of the $\pi\pi$ S-wave states in $\Dz \to \KS \pim \pip$ uses a K-matrix 
formalism~\cite{ref:Kmatrix,ref:aitchison} to account 
for the non-trivial dynamics due to the presence of broad and overlapping resonances. The two models have been obtained
using a high statistics flavor tagged \Dz sample ($D^{\ast +} \rightarrow \Dz \pip_s$) selected 
from $e^+ e^- \ra c \bar c$ events recorded by \babar.

\section{THE \babar\ DETECTOR AND DATASET}
\label{sec:babar}

The analysis is based on a sample of 227 million $B\bar B$ pairs collected by the \babar\ detector at the SLAC PEP-II
 $e^+ e^-$ asymmetric-energy storage ring. \babar\ is a solenoidal detector optimized for the asymmetric-energy beams
at PEP-II and is described in ~\cite{ref:babar}. We summarize briefly
the components that are crucial to this analysis.
Charged-particle tracking is provided by a five-layer silicon
vertex tracker (SVT) and a 40-layer drift chamber (DCH).
In addition to providing precise spatial hits for tracking, the SVT and DCH
also measure the specific ionization ($dE/dx$), which is used for particle
identification of low-momentum charged particles. At higher momenta ($p>0.7$~\gevc)
pions and kaons are identified by Cherenkov radiation detected in a ring-imaging
device (DIRC). The typical separation between pions and kaons varies from 8$\sigma$
at 2~\gevc to 2.5$\sigma$ at 4~\gevc. 
The position and energy of neutral clusters (photons) are
measured with an electromagnetic calorimeter (EMC) consisting of 6580 thallium-doped CsI crystals.
These systems are mounted inside a 1.5-T solenoidal super-conducting magnet.

\section{EVENT SELECTION}
\label{sec:selection}
\begin{figure}[!htb]
\begin{center}
\includegraphics[height=8cm]{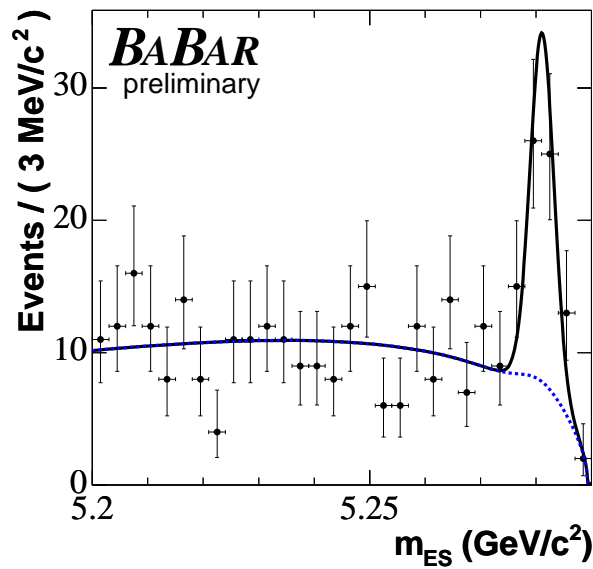}
\caption{$\Bm\to \Dztilde \Kstarm$ \mes distribution after all selection
criteria are applied. The curves represent the fit projections for signal plus background (solid) and background (dotted).
}
\label{fig:variables}
\end{center}
\end{figure}

We reconstruct the decays $\Bm\to \Dztilde \Kstarm$ with $\Dztilde \to \KS \pim \pip$,
$\Kstarm\to \KS\pim$ and $\KS\to\pim\pip$.
The \KS candidates are formed from oppositely charged pions with a reconstructed
invariant mass within 9~\mevcc of the nominal \KS mass~\cite{ref:pdg2004}.
The two pions are constrained to originate from the same point. 
The \Dztilde candidates are selected by combining mass constrained \KS candidates with two oppositely
charged pions having an invariant mass within 12~\mevcc of the nominal \Dztilde mass~\cite{ref:pdg2004}.
The \Dztilde candidates are mass and vertex constrained.
The \Kstarm candidates are selected from combinations of a \KS with a 
negative charged pion with an invariant mass within 55~\mevcc of the nominal \Kstarm mass~\cite{ref:pdg2004}.
The cosine of the angle between the direction transverse to the beam connecting the \Dztilde or \Kstarm
and the \KS decay points (transverse flight direction), 
and the \KS transverse momentum vector is required to be larger than 0.99.
Since the \Kstarm in $\Bm \to \Dztilde \Kstarm$ is polarized, we require $|\cos\theta_H|\ge 0.35$, where
$\theta_H$ is the angle in the \Kstarm rest frame between the daughter pion and the parent \B momentum.
The distribution of $\cos\theta_H$ is proportional to $\cos^2\theta_H$ for the $\Bm\to \Dztilde \Kstarm$ signal
and is roughly flat for the $e^+e^-\to q\bar{q}$ ($q=u,d,s,c$) continuum background.
The \Bm candidates are reconstructed by combining a \Dztilde candidate with a \Kstarm candidate. 
We select the \B mesons by using the beam-energy substituted mass, 
$\mes= \sqrt{(E^{*2}_i/2+{\bf p_i\cdot p_B})^2/E^2_i-p^2_B}$, and the energy difference $\Delta E=E^*_B-E^*_i/2$,
where the subscripts $i$ and $B$ refer to the initial $e^+e^-$ system and the $B$ candidate, respectively, and
the asterisk denotes the center-of-mass (CM) frame. The resolutions of \mes\ and \DeltaE, evaluated on simulated signal events, are 2.6~\mevcc and 
11~\mev, respectively. We define a selection region through the requirement $|\de|<25$~\mev and $\mes>5.2$~\gevcc.
To suppress the background from continuum events we require $|\cos\theta_T|< 0.8$ where $\theta_T$ is defined
as the angle between the thrust axis of the \B candidate and that of the rest of the event. 
After all the cuts are applied the average number of candidates per
event is 1.06. We select one \B candidate per
event by taking the one that has the minimum value of a $\chi^2$ built with the \Dztilde and \Kstar masses,
resolutions and intrinsic width for the case of the \Kstar.
The $\Bm\to \Dztilde \Kstarm$ reconstruction efficiency is $11.1\pm 0.5$\% for simulated events. The reconstruction
purity in the signal region $\mes>5.272$~\gevcc is estimated to be 46\%.
Figure~\ref{fig:variables} shows the \mes distribution after all the selection criteria are applied.

\section{THE $\Dz \to \KS\pim\pip$  DECAY MODEL}
\label{sec:dalitzModel}

\begin{figure}[!t]
\begin{center}
\begin{tabular} {rr}  
\includegraphics[height=6cm]{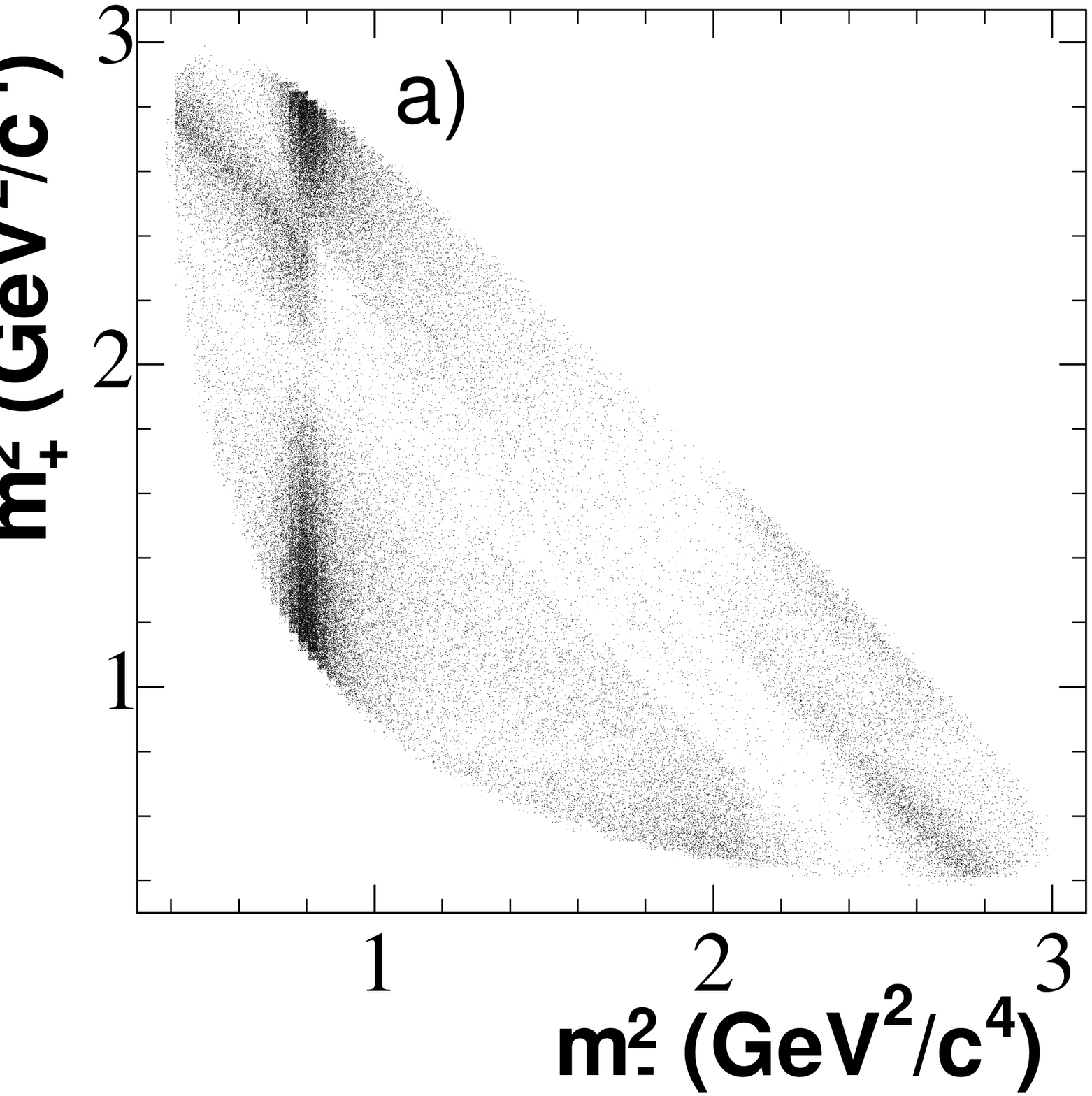} &
\includegraphics[height=6cm]{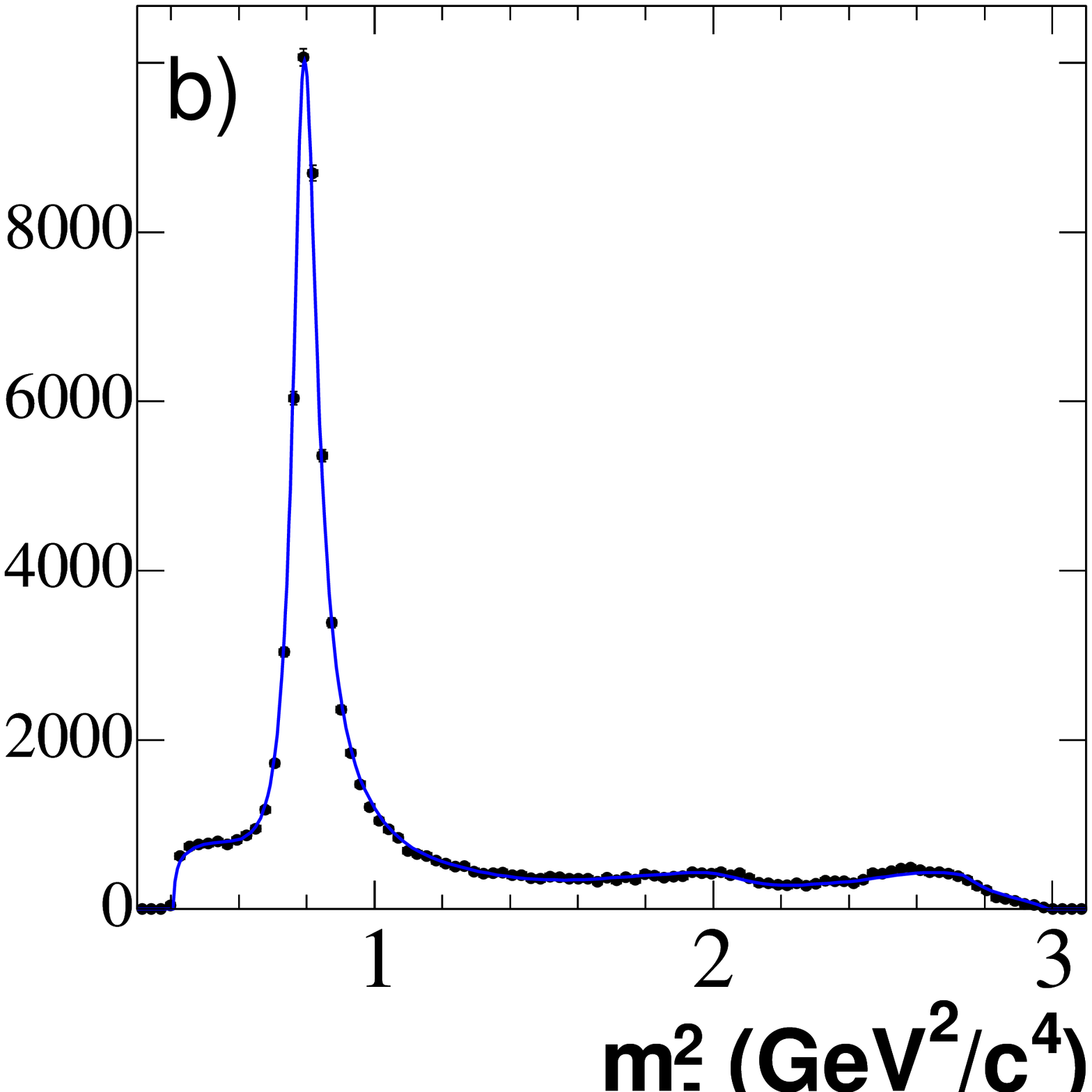} \\
\includegraphics[height=6cm]{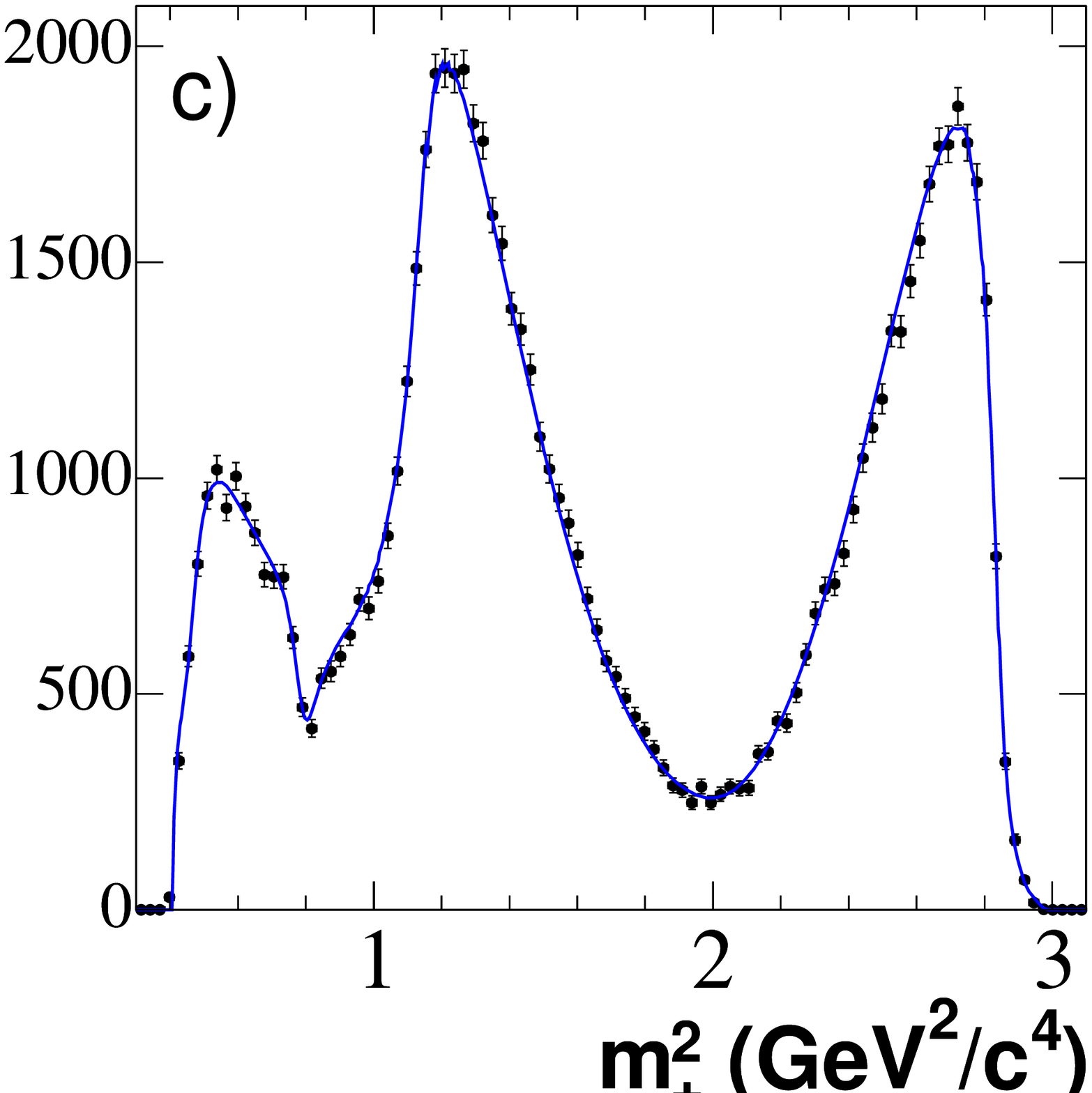} &
\includegraphics[height=6cm]{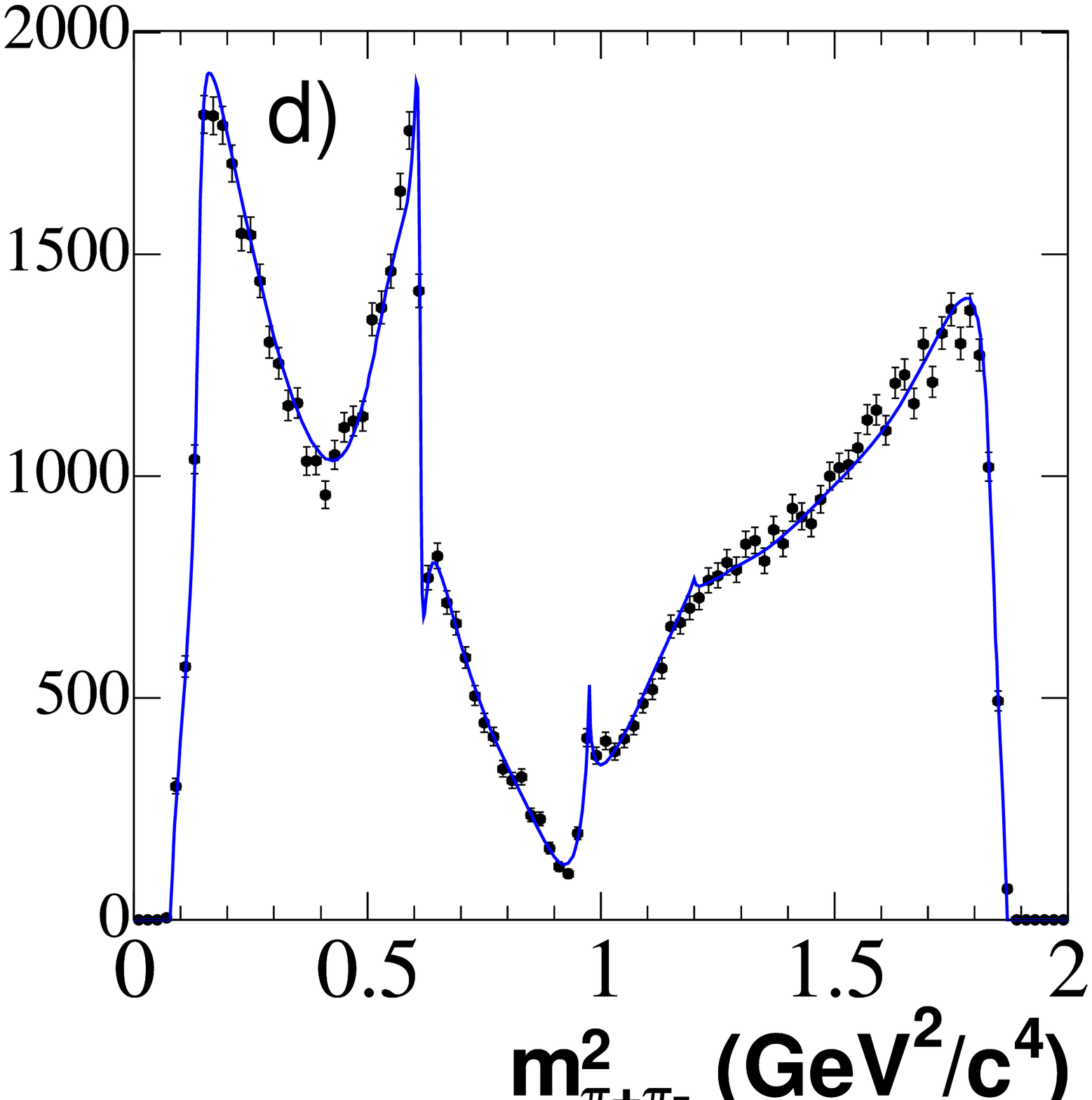} \\
\end{tabular}   
\caption{(a) The $\Dz \to \KS \pim \pip$ Dalitz distribution from $\Dstarp \to \Dz \pip$ events, and projections
on (b) $m^2_-$, (c) $m^2_+$, and (d) $m^2_{\pip\pim}$. $\Dzb \to \KS \pi^+ \pi^-$ from $\Dstarm \to \Dzb \pim$ events 
are also included. 
The curves are the $\pi\pi$ S-wave K-matrix model fit projections.}
\label{fig:dalmkspidcs}
\end{center}
\end{figure}

The $\Dz \to \KS \pim \pip$ decay amplitude ${\cal A}_D(m^2_-,m^2_+)$ is determined from an unbinned maximum-likelihood Dalitz fit to
the Dalitz plot distribution of a high-purity (97\%) sample of 81496 $\Dstarp\to\Dz\pip$ decays reconstructed in 91.5 \invfb of data, 
shown in Fig.~\ref{fig:dalmkspidcs}(a). We use two different models to describe ${\cal A}_D(m^2_-,m^2_+)$.

The first (Breit-Wigner) model is the same as used for our previously 
reported measurement of $\gamma$ on $\Bm \to \dodstartilde \Km$, $\Dztilde \to \KS \pim \pip$ decays~\cite{ref:babar_dalitz}.
Here, the decay amplitude is expressed as a sum of 
two-body decay-matrix elements (subscript $r$) and a non-resonant (subscript NR) contribution,
\bea
{\cal A}_D(m^2_-,m^2_+) = \Sigma_r a_r e^{i \phi_r} {\cal A}_r(m^2_-,m^2_+) + a_{\rm NR} e^{i \phi_{\rm NR}}~,\nn
\eea
where each term is parameterized with an amplitude $a_r$ and a phase $\phi_r$.
The function ${\cal A}_r(m^2_-,m^2_+)$ is the Lorentz-invariant expression for the matrix element of a \Dz meson 
decaying into $\KS \pim \pip$ through an intermediate resonance $r$, parameterized as a
function of position in the Dalitz plane. 
For $r=\rho(770)$ and $\rho(1450)$ we use the functional form 
suggested in Ref.~\cite{ref:gounarissakurai}, while the remaining resonances 
are parameterized by a spin-dependent relativistic Breit-Wigner distribution~\cite{ref:pdg2004}. 
The model consists of 13 resonances leading to 16 two-body decay amplitudes and phases 
(see Table~I in Ref.~\cite{ref:babar_dalitz}), plus the non-resonant contribution, and accounts
for efficiency variations across the Dalitz plane and the small background contribution (with uniform Dalitz shape).
All the  resonances considered in this model are well established except for the two 
scalar $\pi\pi$  resonances, $\sigma$ and $\sigma'$, whose
masses and widths are obtained from our sample.
The $\sigma$ and $\sigma'$ resonances are introduced in order to obtain a better fit to the data, but
we consider in the evaluation of the systematic errors the possibility that they do
not actually exist. We estimate the goodness of fit through a two-dimensional $\chi^2$ test 
and obtain $\chi^2= 3824$ for $3054-32$ degrees of freedom.
This model is the one used as nominal in this analysis.

The second ($\pi\pi$ S-wave K-matrix) model uses the K-matrix formalism~\cite{ref:Kmatrix,ref:aitchison} to parameterize 
the S-wave component of the $\pi\pi$ system in $\Dz \to \KS \pim \pip$. The K-matrix approach can be 
applied to the case of resonance production in multi-body decays when the two-body system in the final state is isolated, 
and the two particles do not interact simultaneously with the rest of the final state
in the production process. In addition, it provides 
a direct way of imposing the unitarity constraint that is not guaranteed in the 
case of the Breit-Wigner model. Therefore, the K-matrix method is suited to the study of broad and overlapping 
resonances in multi-channel decays, solving the main limitation of the Breit-Wigner model to parameterize
the $\pi\pi$ S-wave states in $\Dz \to \KS \pim \pip$~\cite{ref:RevScalarMesons}, and avoiding 
the need to introduce the two $\sigma$ scalars. 

The Dalitz amplitude ${\cal A}_D(m_-^2,m_+^2)$ is written in this case as a sum of two-body decay matrix elements
for the spin-1, spin-2 and $K\pi$ spin-0 resonances (as in the Breit-Wigner model), and the $\pi\pi$ spin-0 piece 
denoted as $F_1$ is written in terms of the K-matrix. We have
\bea
{\cal A}_D(m^2_-,m^2_+) = F_1(s) + \Sigma_{r\neq\pi\pi\,{\rm S-wave}} a_r e^{i \phi_r} {\cal A}_r(m^2_-,m^2_+)~,
\label{eq:ampli} 
\eea
where $F_1(s)$ is the contribution of $\pi\pi$ S-wave states,
\begin{eqnarray}
F_1(s) = \sum_{j} \left[I - i K(s) \rho(s)\right]^{-1}_{1j} P_j(s)~.
\label{eq:flvector}
\end{eqnarray}
Here, $s$ is the squared mass of the $\pi\pi$ system $(m^2_{\pip\pim})$, $I$ is the identity matrix, 
$K$ is the matrix describing the $S$-wave scattering process, $\rho$ is the phase-space matrix, and 
$P$ is the initial production vector~\cite{ref:aitchison},
\begin{eqnarray}
P_{j}(s)= \sum_{\alpha} \frac{\beta_{\alpha} g^{\alpha}_j}{m^2_{\alpha}-s} 
+ f^{\rm prod}_{1j}\frac{1-s^{\rm scatt}_0}{s-s^{\rm scatt}_0} ~.
\label{eq:pvector}
\end{eqnarray}
The index $j$ represents the $j^{\rm th}$ channel ($1=\pi\pi$, $2=K\overline{K}$, $3=$multi-meson\footnote{Multi-meson channel refers
to a final state with four pions.}, 
$4=\eta\eta$, $5=\eta\eta'$~\cite{ref:AS}). The K-matrix parameters are obtained 
from Ref.~\cite{ref:AS} from a global fit of the available 
$\pi\pi$ scattering data from threshold up to $1900\mevcc$. The K-matrix parameterization is
\begin{eqnarray}
K_{ij}(s)=\left\{ \sum_{\alpha} \frac{g^{\alpha}_i g^{\alpha}_j}{m^2_{\alpha}-s} 
+ f^{\rm scatt}_{ij}\frac{1.0-s^{\rm scatt}_0}{s-s^{\rm scatt}_0} \right\}
\frac{(1-s_{A0})}{(s-s_{A0})}(s-s_A m^2_{\pi}/2),
\label{eq:kmatrix}
\end{eqnarray}
where $g^{\alpha}_i$ is the coupling constant of the K-matrix pole $m_{\alpha}$ to the $i^{\rm th}$
channel. 
The parameters $f^{\rm scatt}_{ij}$ and $s^{\rm scatt}_0$ describe the slowly-varying part of the
K-matrix element. The Adler zero factor 
$(1-s_{A0}) (s-s_A m^2_{\pi}/2) / (s-s_{A0}) $~\cite{ref:FOCUS}
suppresses false kinematical singularity at $s=0$ in the physical region near the $\pi\pi$ threshold~\cite{ref:Adler}.
Note that the production vector has the same poles as the K-matrix, otherwise the $F_1$ vector would vanish (diverge)
at the K-matrix (P-vector) poles.
The parameter values used in this analysis are listed in Table~\ref{tbl:AS}~\cite{ref:ASprivate}. 
The parameters $f^{\rm scatt}_{ij}$, for $i \ne 1$, are all set to zero since they are not related 
to the $\pi\pi$ scattering process.

\begin{table}[!tbp]
\caption{\label{tbl:AS} K-matrix parameters as obtained from Refs.~\cite{ref:AS,ref:ASprivate}. 
Pole masses ($m_{\alpha}$) and coupling constants ($g_i^\alpha$) are given in \gevcc, $s^{\rm scatt}_0$ and $s_{A0}$ are in \gevtwo.}
\begin{center}
\begin{tabular}{c|ccccc}
\hline
$m_{\alpha}$ & $g_{\pipi}^\alpha$ & $~g_{K\overline{K}}^\alpha$ & $~g_{4\pi}^\alpha$ & $~g_{\eta\eta}^\alpha$ & $~g_{\eta\eta^{\prime}}^\alpha$  \\
\hline
     $0.651$  &   $0.229$  &  $          -0.554$  &  $\phantom{-}0.000$  &  $          -0.399$  &  $          -0.346$ \\
     $1.204$  &   $0.941$  &  $\phantom{-}0.551$  &  $\phantom{-}0.000$  &  $\phantom{-}0.391$  &  $\phantom{-}0.315$ \\
     $1.558$  &   $0.369$  &  $\phantom{-}0.239$  &  $\phantom{-}0.556$  &  $\phantom{-}0.183$  &  $\phantom{-}0.187$ \\
     $1.210$  &   $0.337$  &  $\phantom{-}0.409$  &  $\phantom{-}0.857$  &  $\phantom{-}0.199$  &  $          -0.010$ \\
     $1.822$  &   $0.182$  &  $          -0.176$  &  $          -0.797$  &  $          -0.004$  &  $\phantom{-}0.224$ \\
\hline \hline
              &    $f^{\rm scatt}_{11}$  &   $~f^{\rm scatt}_{12}$  &   $~f^{\rm scatt}_{13}$  &   $~f^{\rm scatt}_{14}$  &   $~f^{\rm scatt}_{15}$ \\ 
\cline{2-6}
              &    $0.234$    &   $\phantom{-}0.150$    &   $-0.206$   &     $\phantom{-}0.328$  &   $\phantom{-}0.354$   \\	
\hline \hline
              &  $~s^{\rm scatt}_0$ & $~s_{A0}$ & $~s_{A}$ & & \\
\cline{2-4}
              &   $-3.926$  & $-0.15$ & $\phantom{-}1$ & & \\
\hline
\end{tabular}
\end{center}
\end{table}

The phase space matrix is diagonal, $\rho_{ij}(s) = \delta_{ij} \rho_{i}(s)$, where 
\begin{eqnarray}
\rho_{i}(s) = \sqrt{1-\frac{  (m_{1i} + m_{2i})^2  }{s}}~,
\label{eq:rhops}
\end{eqnarray}
with $m_{1i}$ ($m_{2i}$) denoting the mass of the first (second) final state particle of the $i^{th}$ channel.
The normalization is such that $\rho_i \rightarrow 1$ as $s \rightarrow \infty$.
We use an analytic continuation of the $\rho_i$ functions below threshold.
The expression of the multi-meson state phase space is written as~\cite{ref:AS}

\begin{eqnarray}
\rho_3(s) &= \left\{\begin{array}{ll}
                    \rho_{31} & s<1~\gev^{2} \\
                    \rho_{32} & s>1~\gev^{2}
                    \end{array}\right\},
\end{eqnarray}
where
\bea
\rho_{31}(s)&=&\rho_0\int\int\frac{ds_1}{\pi}\frac{ds_2}{\pi}
\frac{M^2 \Gamma(s_1)\Gamma(s_2)\sqrt{(s+s_1-s_2)^2-4ss_1}}
{s[(M^2-s_1)^2+M^2\Gamma^2(s_1)]\,[(M^2-s_2)^2+M^2\Gamma^2(s_2)]},\nn \\
\rho_{32}(s)&=&\frac{s-16m^2_{\pi}}{s}~.
\label{eq:rho3ps}
\eea
Here $s_1$ and $s_2$ are the squared invariant mass of the two dipion systems,
$M$ is the $\rho$ meson mass, and $\Gamma(s)$ is the energy-dependent width. The factor $\rho_0$ provides
the continuity of $\rho_3(s)$ at $s=1~\gev^2$.  Energy conservation in the dipion system 
must be satisfied when calculating the integral. 
This complicated expression reveals the fact that the $\rho$ meson
has an intrinsic width.  If one sets $\Gamma(s) = \delta(s)$, where $\delta$ is the
Dirac $\delta$ function, the usual two-body phase space factor is obtained.

Table~\ref{tab:fitreso-likelihood} summarizes the values of the P-vector free parameters
$\beta_{\alpha}$ and $f^{\rm prod}_{11}$ (we are describing only $\pi\pi$ channel), together with
the spin-1, spin-2, and $K\pi$ spin-0 amplitudes as in the Breit-Wigner model. The third and fifth poles 
are not included since they are far beyond our $\pi\pi$ kinematic range. 
Figures~\ref{fig:dalmkspidcs}(b,c,d) show the fit projections overlaid with the data distributions. 
There is no overall improvement in the two-dimensional $\chi^2$ test compared to the Breit-Wigner model~\cite{ref:babar_dalitz}
since it is dominated by the P-wave components, which are identical between the two models. The total
fit fraction is slightly changed from 1.24 to 1.16. Nevertheless, it should be emphasized that the 
main advantage of using a K-matrix parameterization rather than a sum of two-body amplitudes to describe the $\pi\pi$ S-wave
is that it provides a more adequate description of the complex dynamics in the presence of overlapping and many
channel resonances.

\begin{table}[!h]
\caption{Complex amplitudes $a_r e^{i\phi_r}$ and fit fractions of the different components 
($\KS\pim$ and $\KS\pip$ resonances, and $\pip\pim$ poles) obtained from the fit of 
the $\Dz \to \KS\pim\pip$ Dalitz distribution from $\Dstarp \to \Dz \pip$ events. 
Errors are statistical only.
Masses and widths of all resonances are taken from~\cite{ref:pdg2004}, while the pole masses and scattering 
data are from~\cite{ref:AS,ref:ASprivate}.
The fit fraction is defined for the resonance terms ($\pi\pi$ S-wave term) as the integral 
of $a_r^2 |{\cal A}_r(m^2_-,m^2_+)|^2$ ($|F_1(s)|^2$) over the Dalitz plane
divided by the integral of $|{\cal A}_D(m^2_-,m^2_+)|^2$. The sum of fit fractions is $1.16$.}
\label{tab:fitreso-likelihood}
\begin{center}
\begin{tabular}{l|ccc}
\hline
    Component  &  $\re\{a_r e^{i\phi_r}\}$ &  $\im\{a_r e^{i\phi_r}\}$ & Fit fraction (\%) \\ 
\hline \hline
$K^{*}(892)^-$        &    $-1.159 \pm 0.022$    &  $1.361 \pm 0.020$   &  $58.9$    \\ 
$K^{*}_0(1430)^-$     &    $2.482 \pm 0.075$     &  $-0.653 \pm 0.073$   &  $9.1$    \\ 
$K^{*}_2(1430)^-$     &    $0.852 \pm 0.042  $   &  $-0.729 \pm 0.051$   &  $3.1$    \\ 
$K^{*}(1410)^-$       &    $-0.402 \pm 0.076$    &  $0.050 \pm 0.072$   &  $0.2$    \\ 
$K^{*}(1680)^-$       &    $-1.00 \pm 0.29$   &  $1.69\pm 0.28$   &  $1.4$    \\ 
\hline
$K^{*}(892)^+$     &    $0.133 \pm 0.008$   &  $-0.132\pm 0.007$  &  $0.7$     \\ 
$K^{*}_0(1430)^+$  &    $0.375 \pm 0.060$   &  $-0.143\pm0.066$  &  $0.2$    \\ 
$K^{*}_2(1430)^+$  &    $0.088 \pm 0.037$   &  $-0.057\pm0.038$  &  $0.0$    \\ 
\hline 
$\rho(770)$         &    1 (fixed)  &    0 (fixed)  &   $22.3$    \\ 
$\omega(782)$       &    $-0.0182\pm 0.0019$   &  $0.0367\pm0.0014$   &     $0.6$     \\ 
$f_2(1270) $        &    $0.787\pm 0.039$    &  $-0.397\pm0.049$   &     $2.7$     \\ 
$\rho(1450)$        &    $0.405\pm 0.079$   &  $-0.458\pm0.116$   &     $0.3$    \\ 
\hline \hline
$\beta_1$           &    $-3.78 \pm 0.13$  &  $1.23\pm0.16$   &     $-$     \\ 
$\beta_2$           &    $9.55 \pm 0.20$   &  $3.43\pm0.40$   &     $-$    \\ 
$\beta_4$           &    $12.97 \pm 0.67$   &  $1.27\pm0.66$   &     $-$    \\ 
$f_{11}^{\rm prod}$ &    $-10.22 \pm 0.32$  &  $-6.35\pm0.39$   &     $-$     \\ 
sum of $\pip\pim$ S-wave & &                 &     $16.2$ \\
\hline
\end{tabular}
\end{center}
\end{table}

\section{CP ANALYSIS}
\label{sec:cpfit}

We perform an unbinned extended maximum-likelihood fit to the $\Bm\to \Dztilde \Kstarm$ sample
to extract the \CP-violating parameters $\xsmp$, $\ysmp$, and $\rstwo$ along with the signal 
and background yields. 
The $\xspm$ and $\yspm$ variables are more suitable fit parameters than 
$\kappa\rs$, \deltas and \g because they are better behaved near the origin, especially in low-statistics samples.
The fit uses \mes\ and the same Fisher discriminant \fis\ as used in~\cite{ref:babar_dalitz} to distinguish events from
\BB production and continuum background.
The Fisher is a linear combination of four topological variables: $L_0=\sum_{i} p_i^*$, $L_2 =\sum_{i} p_i^*  |\cos \theta^*_i|^2$, 
and the absolute values of the cosine of the CM polar angles of the \B candidate
momentum and thrust direction. Here, $p_i^*$ and $\theta_i^*$ are the
CM momentum and the angle with respect to
the \B candidate thrust axis of the remaining tracks and clusters in the event.
The likelihood for candidate $j$ is obtained by summing the product of the event yield $N_c$, 
the probability density functions (PDF's) for the kinematic and event shape variables ${\cal P}_{c}$,
and  the Dalitz distributions ${\cal P}_{c}^{\rm Dalitz}$, over the
signal and background components $c$. 
The overall likelihood function is
\bea
{\cal L} = \exp\left(-\sum_c N_c\right) \prod_j \sum_c N_c{\cal P}_c(\vec{\xi}_j){\cal P}^{\rm Dalitz}_c(\vec{\eta}_j)~,\nn
\eea
where $\vec{\xi}_j = \{\mes,\fis\}_j$, $\vec{\eta}_j = (m_-^2,m_+^2)_j$, and ${\cal P}_c(\vec{\xi}) = {\cal P}_c(\mes) {\cal P}_c(\fis)$.
The components in the fit are signal, continuum background, and \BB background.
For signal events, ${\cal P}^{\rm Dalitz}_c(\vec{\eta})$ is given by $\Gamma_{\mp}(\vec{\eta})$ 
corrected by the efficiency variations, where $\Gamma_{\mp}(\vec{\eta})$ is given by Eq.~(\ref{eq:ampgen2}). 
The \mes (\fis) distribution for signal events is described by a Gaussian (double-Gaussian) function distribution 
whose parameters are determined from a fit to the same $\Bm \to \Dz \pim$ high-statistics control sample as in our previous 
analysis~\cite{ref:babar_dalitz}.

\subsection{Background composition}

The event yields $N_c$ for signal, continuum, and \BB components are, respectively, $42\pm8$, $251\pm24$, and $45\pm21$, 
in agreement with our expectation from simulation and measured branching ratios. The dominant background contribution is from the
random combination of a real or fake \Dz meson with a charged track and a \KS\ in continuum events or other \BB decays. 
The continuum background in the \mes distribution is described by a threshold function \cite{ref:argus} whose 
free parameter $\zeta$ is determined from the $\Bm \to \Dz \pim$ control sample. The Fisher PDF for continuum background 
is determined using $\Bm \to \Dz \pim$ events from the \mes\ sideband region.
The shape of the background \mes distribution in generic \BB decays is taken from simulation and
uses a threshold function to describe the combinatorial component plus a Gaussian distribution to parameterize the peaking 
contribution arising from events with a misreconstructed pion and having a topology similar to that of signal events. 
The Gaussian component has a width of $3.8\pm1.4$~\gevcc and its fraction with respect to the total \BB background 
is $0.13\pm0.03$. The Fisher PDF and the mean of the \mes Gaussian for \BB events are assumed to be the same as that for the signal.

An important class of background events arises from continuum and \BB background events where a real \Dz is produced back-to-back 
with a \Kstar in the CM. Depending on the flavor-charge correlation this background can mimic 
either the $\b \to \c$ or the $\b \to \u$ signal component. 
In the likelihood function we take this effect into account with two parameters, the fraction $f_{\Dz}$ of background events with 
a real \Dz  and the parameter $R$, the fraction of background events with a real \Dz associated with an oppositely 
flavored kaon (same charge correlation as the $\b \to \u$ signal component). These fractions have been evaluated
separately from continuum and generic \BB simulated events, and are found to be
$f_{\Dz}^{\qqbar}=0.21\pm0.02$, $f_{\Dz}^{\BB}=0.18\pm0.02$, 
$R^{\qqbar}=0.51\pm0.06$ and $R^{\BB}=0.67\pm0.06$. In addition, to check the reliability of these estimates,
the fraction $f_{\Dz}$ for all background (continuum and \BB) events has been evaluated from data using events 
satisfying $\mes<5.272$~\gevcc after removing the  
requirement on the \Dz mass cut. The measured value of $0.20\pm0.06$ is consistent with the continuum and \BB fractions 
obtained from simulated events. The shapes of the Dalitz plot distributions are parameterized by a third-order
polynomial in $(m_-^2,m_+^2)$ for the combinatorial component (fake \Dz) and as signal \Dz or \Dzb shapes for 
real neutral $D$ mesons. The parameters of the polynomials are extracted from off-resonance data and sidebands for the
continuum component, while we use the simulation for \BB.

A potentially dangerous background originates from signal $\Bm\to \Dztilde \Kstarm$ where the \Dztilde meson is combined
with a random \Kstar or pion from the other \B meson having the opposite charge. Using simulated events the fraction of these wrong sign 
signal events is found to be $(0.43\pm0.05)$\%, and therefore this contribution has been neglected.

\subsection{\CP parameters}

\begin{figure}[hbt]
\begin{center}
\begin{tabular}{ccc}
\includegraphics[height=7cm]{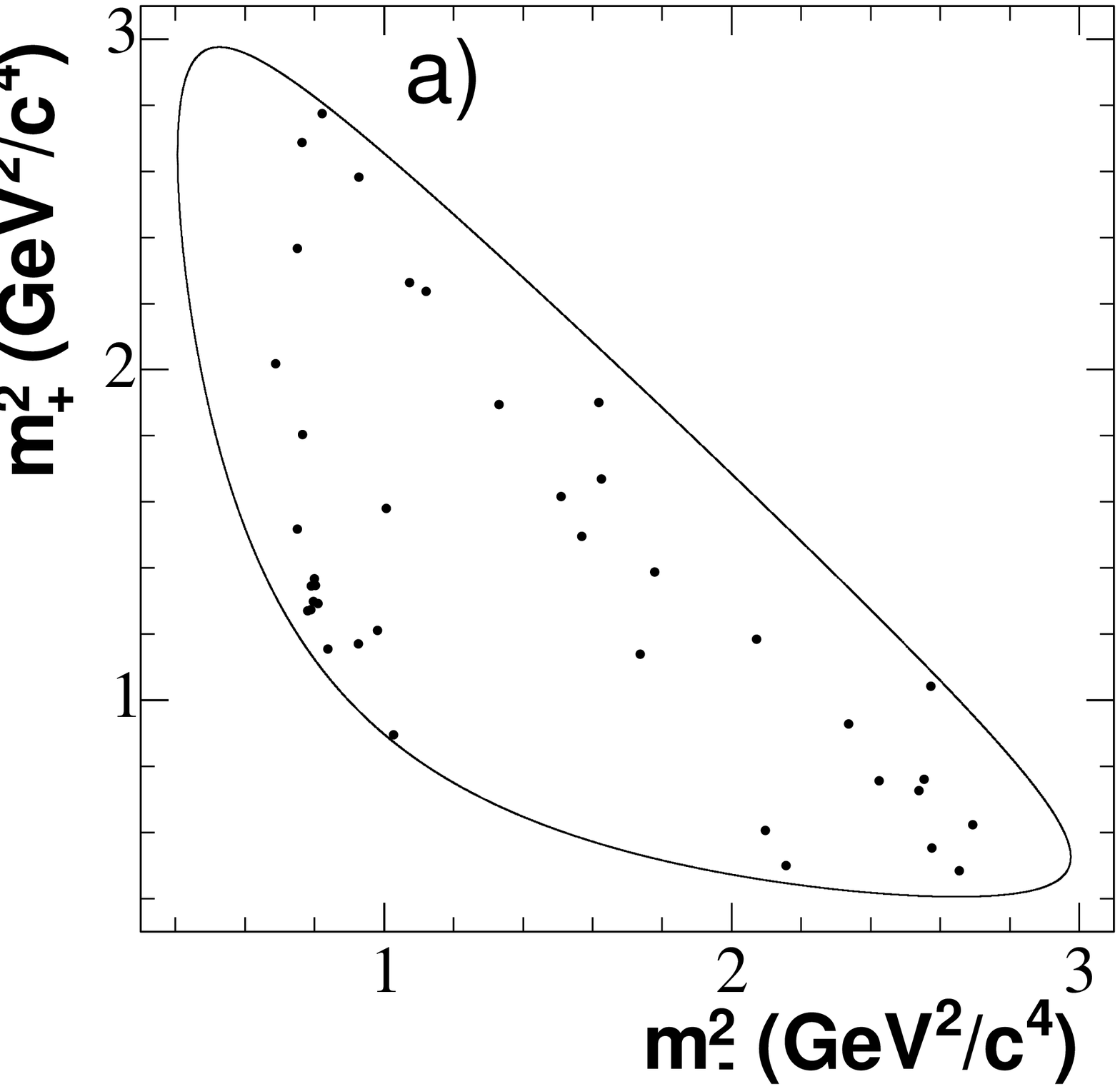} &
\includegraphics[height=7cm]{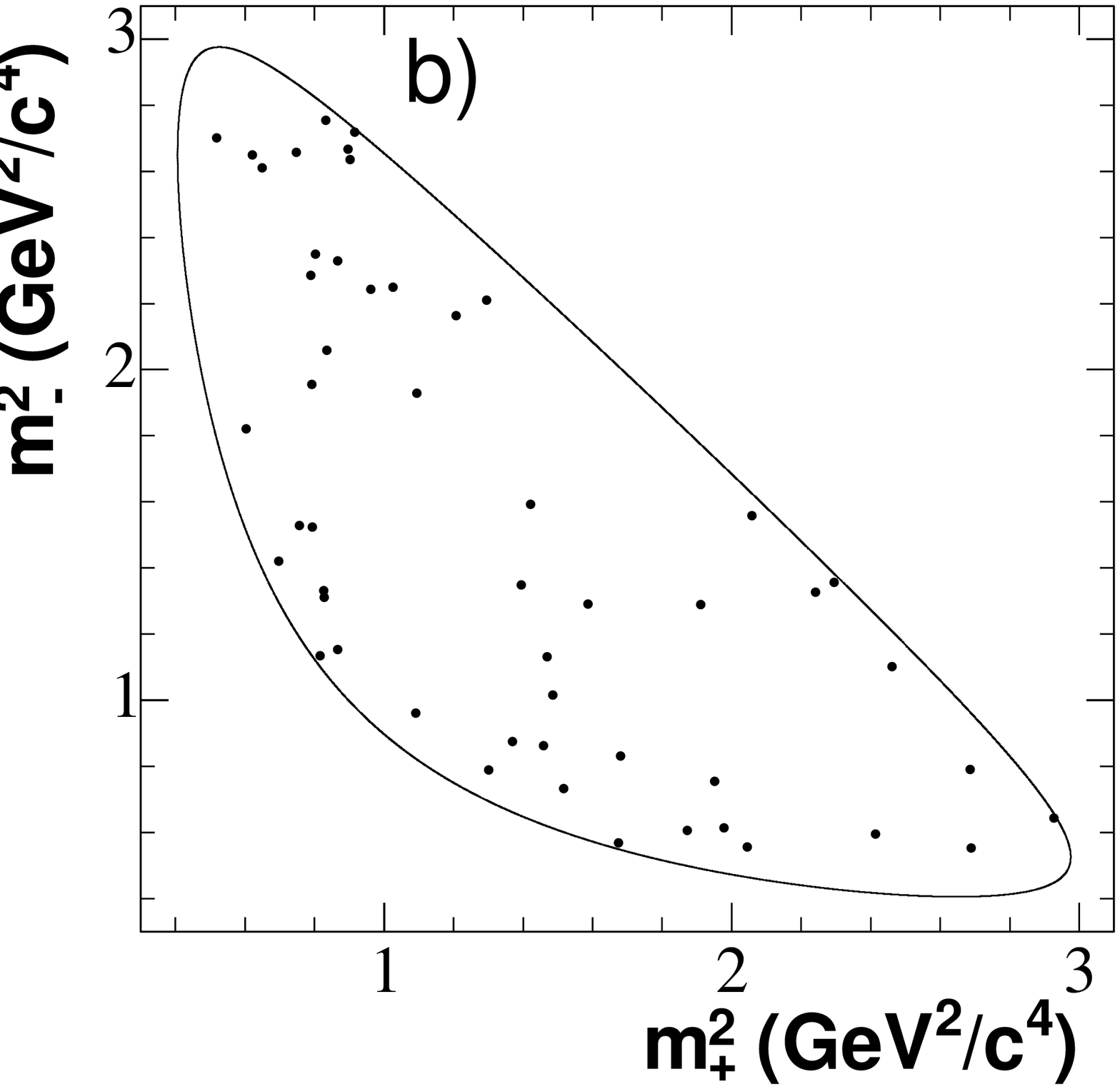}
\end{tabular}
\caption{The $\Dztilde \to \KS\pim\pip$ Dalitz distributions from (a) $\Bm \to \Dztilde \Kstarm$ and (b) $\Bp \to \Dztilde \Kstarp$
events with $\mes>5.272$~\gevcc.
}
\label{fig:DKstDalitz}
\end{center}
\end{figure}

\begin{figure}[hbt]
\begin{center}
\begin{tabular}{ccc}
\includegraphics[height=5.2cm]{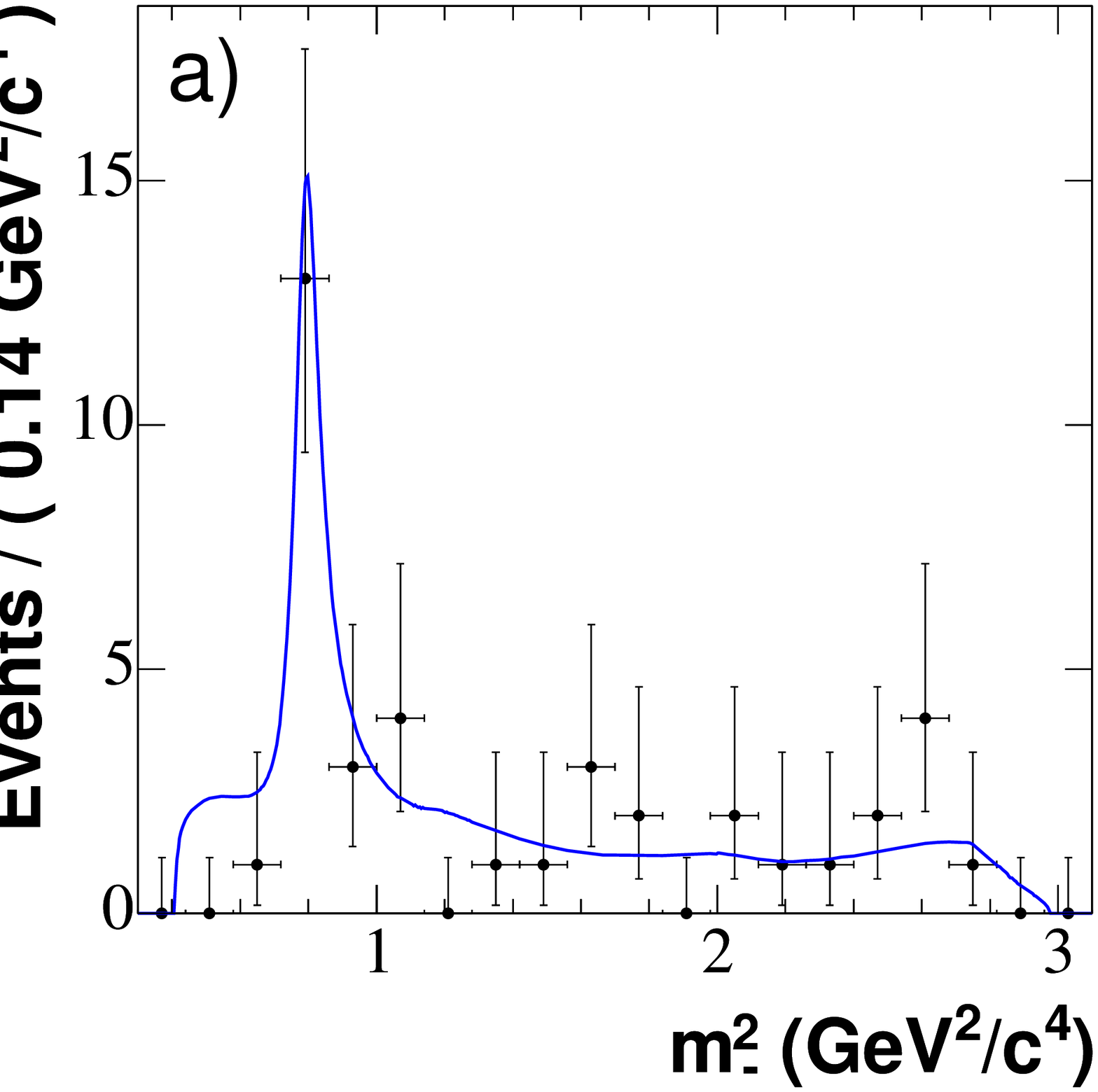} &
\includegraphics[height=5.2cm]{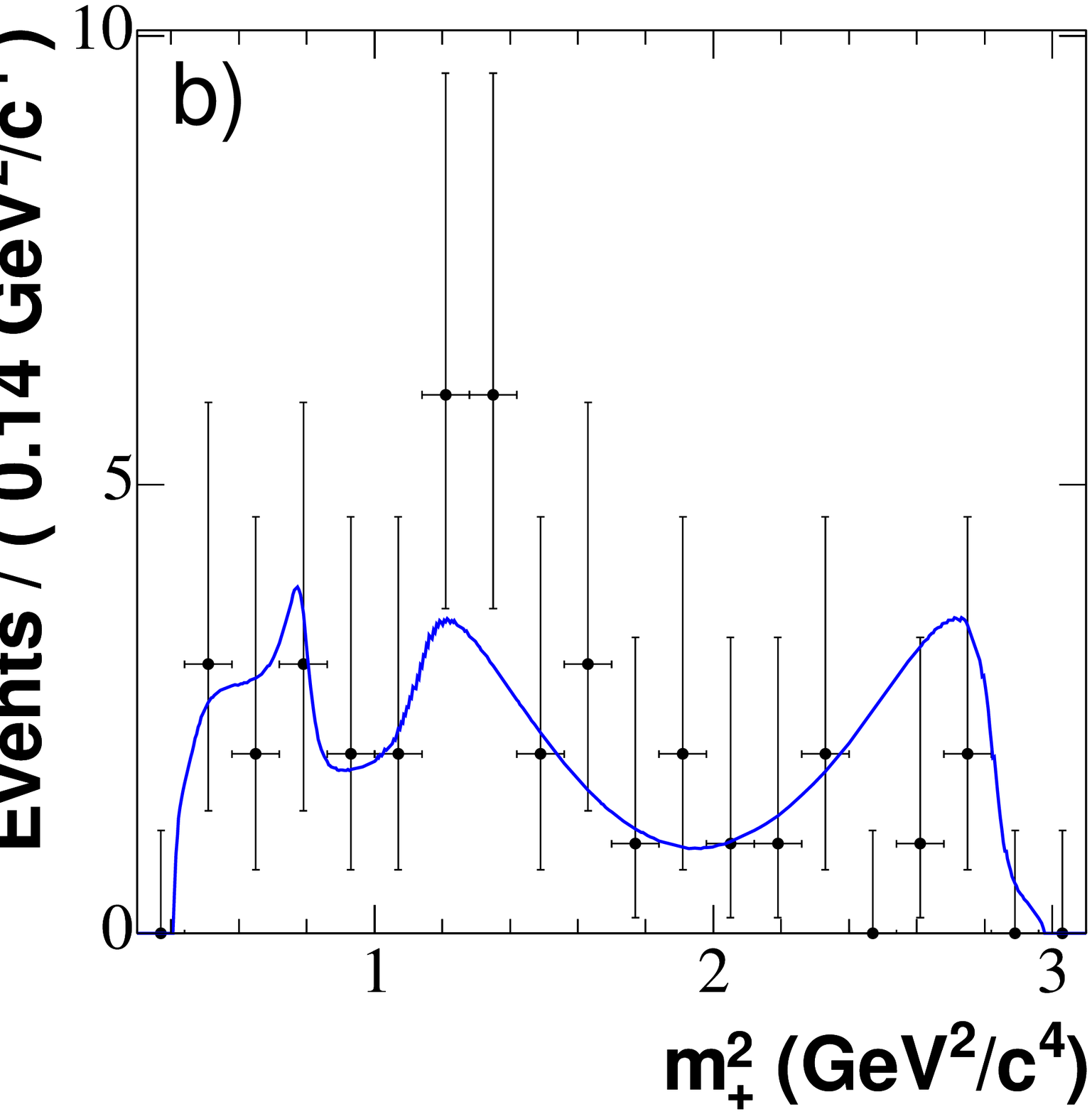} &
\includegraphics[height=5.2cm]{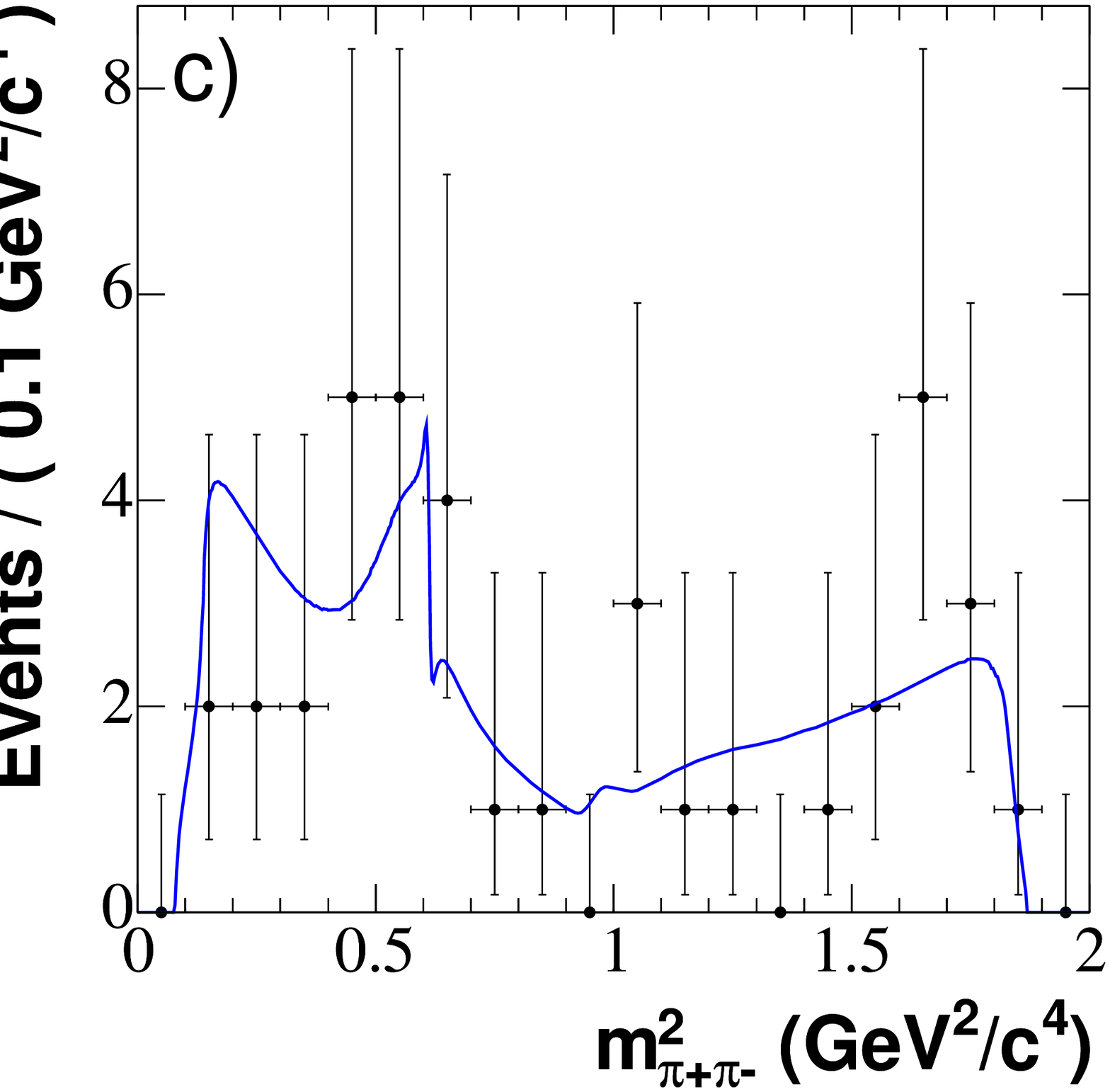} \\
\includegraphics[height=5.2cm]{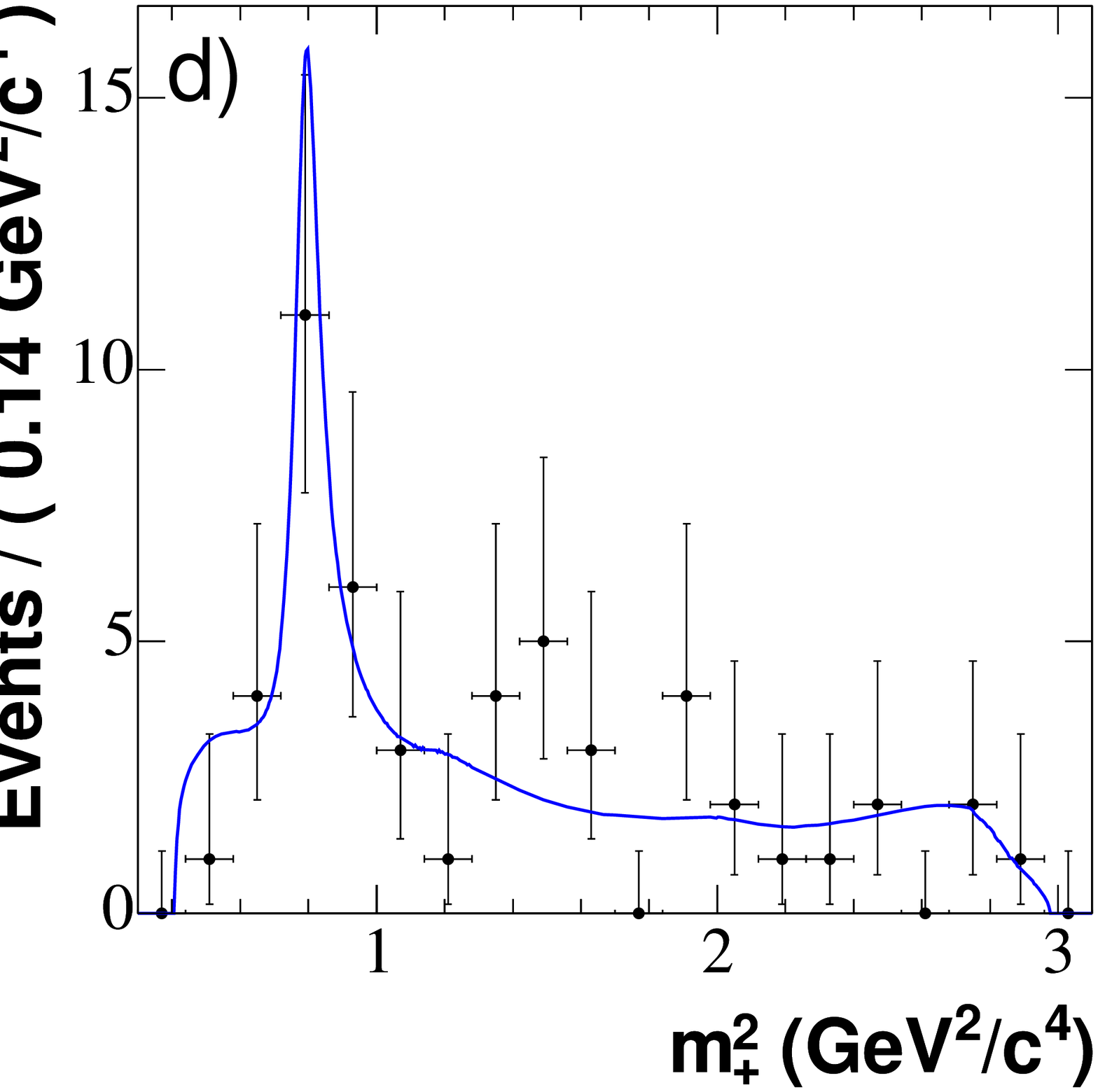} &
\includegraphics[height=5.2cm]{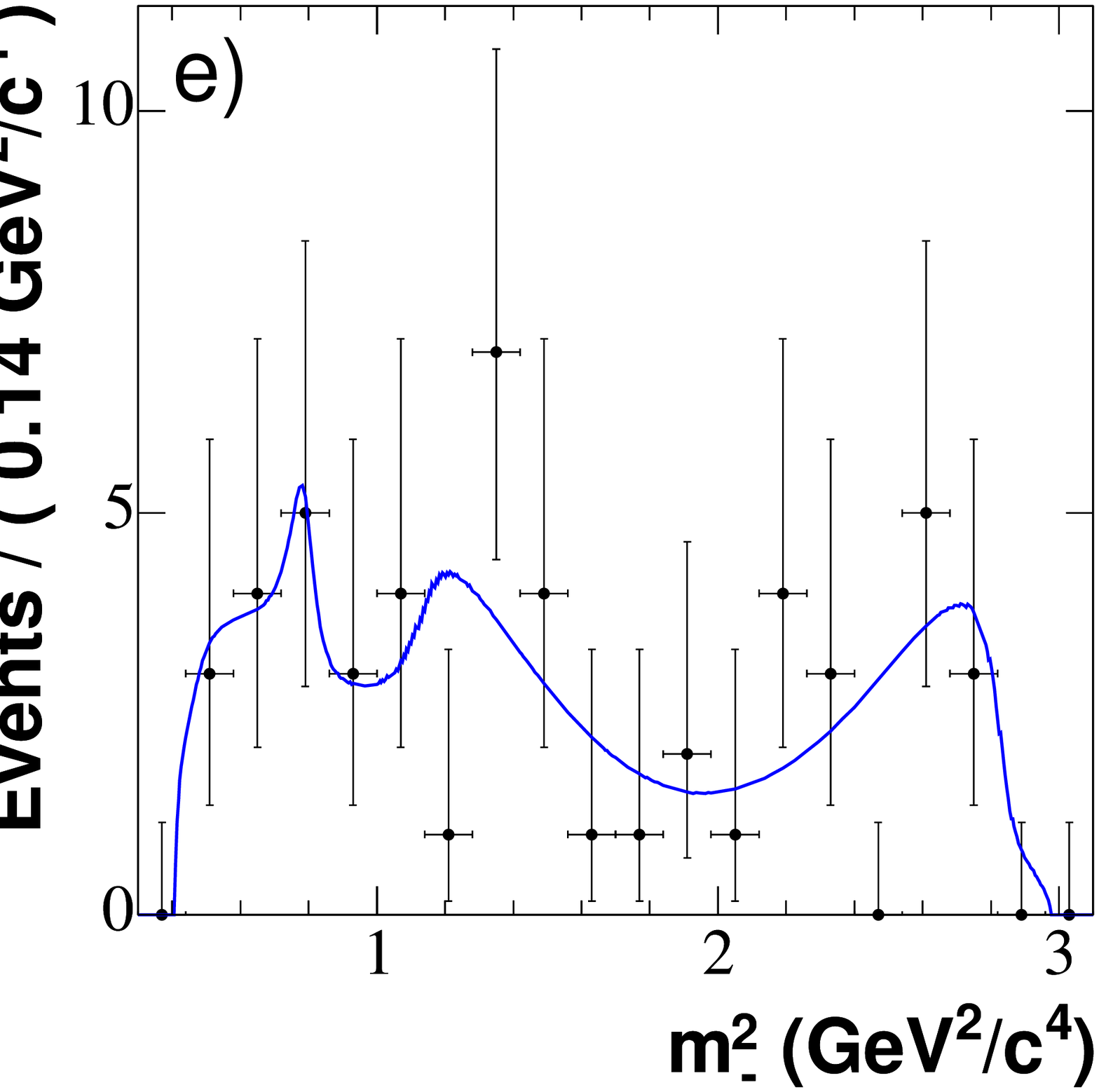} &
\includegraphics[height=5.2cm]{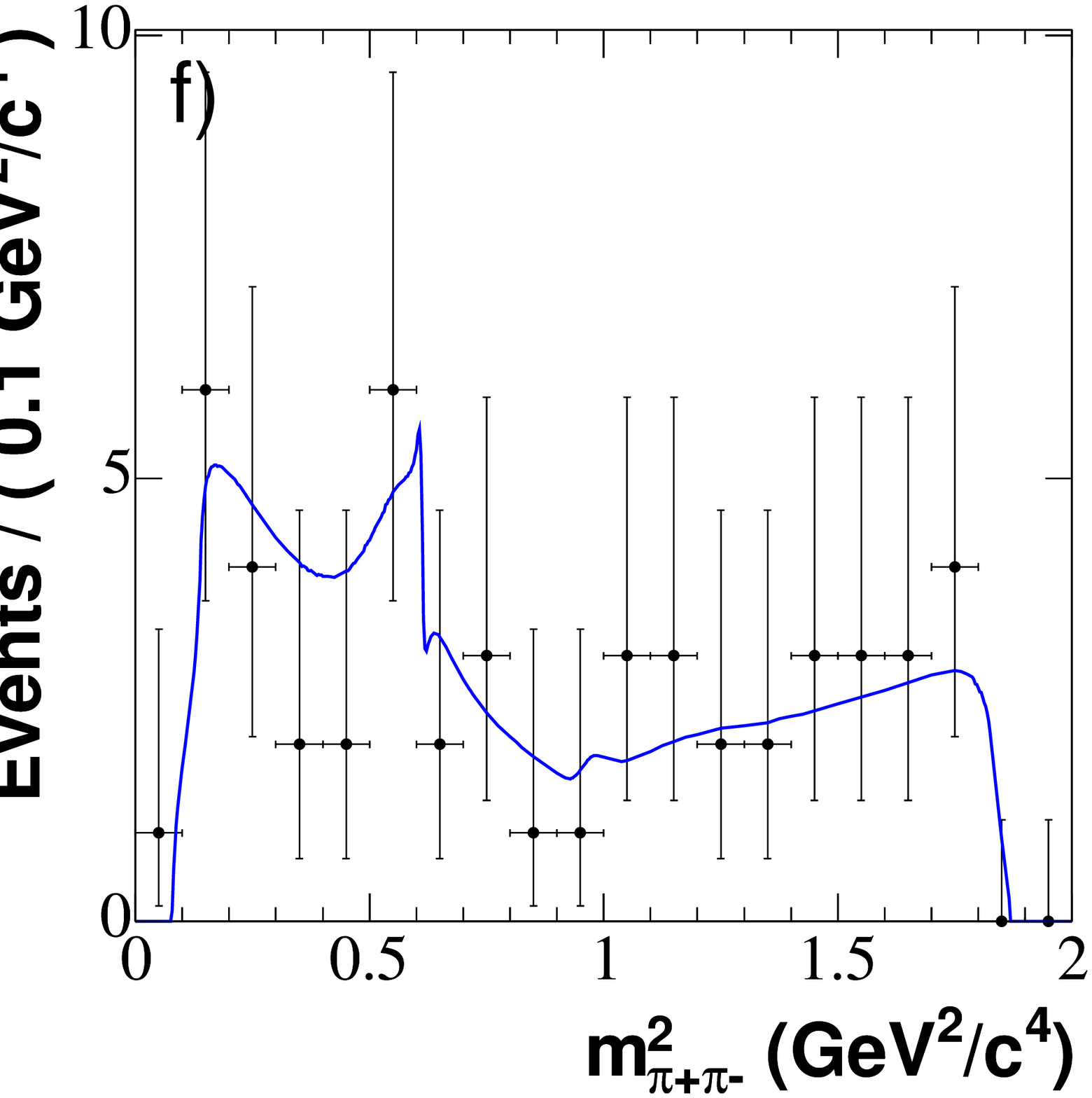} \\
\end{tabular}
\caption{
Projections of the Dalitz distributions on $m^2_-$,  $m^2_+$ and $m^2_{\pip\pim}$ 
from a)b)c) $\Bm \to \Dztilde \Kstarm$ and d)e)f) $\Bp \to \Dztilde \Kstarp$ events with $\mes>5.272$~\gevcc. 
The projections of the fit result are superimposed.
}
\label{fig:DKstprojDalitz}
\end{center}
\end{figure}

\begin{figure}[!htb]
\begin{center}   
 \includegraphics[height=8cm]{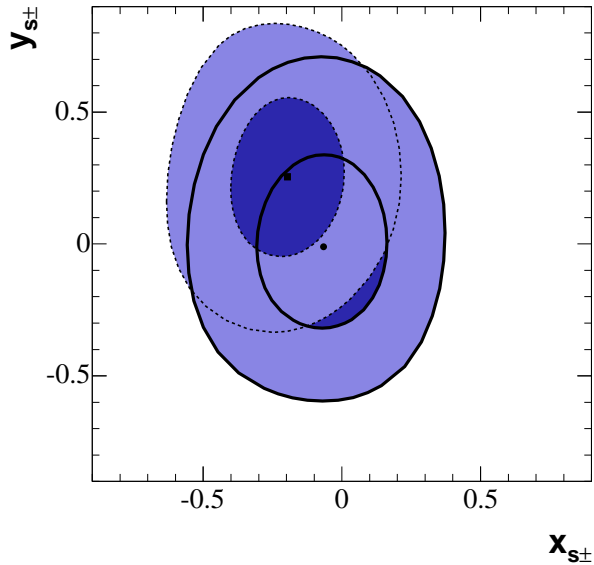}
\caption{Two-dimensional one- (dark) and two- (light) standard deviation regions
(statistical only) in the $(\xs,\ys)$ plane, corresponding to 39.3\% and 86.5\% probability content, separately for  
$\Bm\to \Dztilde \Kstarm$ (thick and solid) and $\Bp\to \Dztilde \Kstarp$ (thin and dotted). The confidence regions 
for \Bp are superimposed over those for \Bm.
}
\label{fig:contours}
\end{center} 
\end{figure}

The Dalitz plot distributions for the $\Dztilde \to \KS\pim\pip$
decay from $\Bm \to \Dztilde \Kstarm$ for events with $\mes>5.272$~\gevcc are shown
in Fig.~\ref{fig:DKstDalitz}. The distributions for \Bm and \Bp candidates are shown separately.
The results for the \CP-violating parameters $\xspm$ and $\yspm$ obtained by fitting those distributions
are summarized in Table~\ref{tab:xyresults}. 
From the same fit we obtain for \rstwo the value $0.05\pm0.11$ (statistical error only).
The only relevant statistical correlations involving the \CP parameters are for the
pairs $(\xsm,\ysm)$ and $(\xsp,\ysp)$, which amount to $-10.1\%$ and $2.2\%$, respectively. The statistical
correlation of \rstwo with \xsm, \ysm, \xsp, and \ysp are $-22.6\%$, $3.7\%$, $-25.1\%$, and $22.0\%$, respectively.
The Dalitz distribution projections on $m^2_-$, $m^2_+$ and $m^2_{\pip\pim}$ for events satisfying $\mes> 5.272$~\gevcc are 
compared to the projection of the fit in Fig.~\ref{fig:DKstprojDalitz},
separately for \Bm and \Bp events.
Figure~\ref{fig:contours} shows the two-dimensional one- (dark) and two- (light) standard deviation regions (statistical only) 
in the $(\xs,\ys)$ plane, corresponding to 39.3\% and 86.5\% probability content, separately for
$\Bm\to \Dztilde \Kstarm$ and $\Bp\to \Dztilde \Kstarp$.
The separation $d$ between the \Bm and \Bp regions in these planes is proportional to the amount 
of direct \CP violation, $d= 2 \kappa \rs |\sin \g|$.

\begin{table}[!t]
\caption{\CP-violating parameters $(\xspm,\yspm)$ obtained from the \CP fit to the $\Bm\to \Dztilde \Kstarm$ sample.
The first error is statistical, the second is the experimental
systematic uncertainty and the third reflects the Dalitz model uncertainty. 
}
\label{tab:xyresults}
\begin{center}
\begin{tabular}{lc}
\hline
\CP parameter & Result \\ 
\hline
$\xsm \equiv  \kappa \rs \cos(\deltas-\g)$ & $          -0.20\pm0.20\pm0.11\pm0.03$ \\
$\ysm \equiv  \kappa \rs \sin(\deltas-\g)$ & $\phantom{-}0.26\pm0.30\pm0.16\pm0.03$ \\
$\xsp \equiv  \kappa \rs \cos(\deltas+\g)$ & $          -0.07\pm0.23\pm0.13\pm0.03$ \\
$\ysp \equiv  \kappa \rs \sin(\deltas+\g)$ & $          -0.01\pm0.32\pm0.18\pm0.05$ \\
\hline
\end{tabular}
\end{center}
\end{table}

\subsection{Experimental systematic errors}

Table~\ref{tab:cartesian-syst} summarizes the break down of the experimental systematic uncertainties.
These include the errors on the \mes and \fis\ PDF parameters for 
signal and background, the uncertainties in the
knowledge of the Dalitz distribution of background events, the efficiency variations across the Dalitz plane, 
and the uncertainty in the fraction of events with a
real \Dz produced in a back-to-back configuration with a negatively-charged kaon.
Less significant systematic uncertainties originate from the imprecise knowledge of the fraction of real \Dz's,
the invariant mass resolution (negligible), tracking efficiency, and the statistical errors in the Dalitz amplitudes and phases from
the fit to the tagged \Dz sample.
We quote as systematic uncertainty, for each effect, the maximum of the difference 
between the bias and square root of the quadratic difference of the statistical error between 
the nominal fit result and the one corresponding to the effect under consideration.
These systematic uncertainties will be reduced with a larger data and 
simulated samples and are not expected to limit the eventual sensitivity of the analysis.

\begin{table}[htpb]
\caption{Summary of the non-negligible contributions to the systematic error on the \CP parameters \xspm and \yspm. 
\label{tab:cartesian-syst}}
\begin{center}
\begin{tabular}{l||c|c|c|c}
\hline
 Source                          & \xsm  & \ysm & \xsp & \ysp  \\   \hline
 \mes, \fis\ shapes              &  0.08    &  0.12   &   0.10  &   0.12    \\ 
 Background Dalitz shape         &  0.04    &  0.09   &   0.04  &   0.09  \\
 Efficiency in the Dalitz plot   &  0.06    &  0.04   &   0.07  &   0.09 \\ 
 Right sign \Dztilde fractions ($R^{\qqbar}$, $R^{\BB}$)
                                 &  0.03    &  0.04   &   0.03  &   0.05 \\ 
 Real \Dztilde\ fractions ($f_{\Dz}^{\qqbar}$, $f_{\Dz}^{\BB}$)             
                                 &  0.03    &  0.03   &   0.03  &   0.04 \\ 
 Tracking efficiency             &  0.01    &  0.01   &   0.01  &   0.01 \\ 
 Dalitz amplitudes and phases    &  0.01    &  0.01   &   0.01  &   0.01  \\ \hline
 Total experimental              &  0.11    &  0.16   &   0.13  &   0.18  \\ \hline 
 Dalitz model (Breit-Wigner model without $\sigma$ scalars)  &  0.03    &  0.03   &   0.03  &   0.05  \\ \hline 
 Dalitz model ($\pi\pi$ S-wave K-matrix model) &  0.01    &  0.01   &   0.01  &   0.01  \\ \hline 
\end{tabular}
\end{center}
\end{table}

\subsection{$\Dztilde \to \KS\pim\pip$ Dalitz model systematic uncertainty}

The largest single contribution to the systematic uncertainties in the \CP parameters 
comes from the choice of the Dalitz model used to describe 
the $\Dz\to\KS\pim\pip$ decay amplitude ${\cal A}_D$. We use the same procedure as in our previous measurement~\cite{ref:babar_dalitz}
to evaluate this uncertainty. We first generate large samples of pseudo-experiments using the nominal (Breit-Wigner) model.
Since the effect on the \CP parameters depends on their generated values, for each pseudo-experiment
we randomly generate the truth values of the Cartesian \CP parameters according to their measured central values and 
statistical errors. We then compare experiment by experiment the values of the \xspm and \yspm obtained from
fits using the nominal model and a set of alternative models. We find that models resulting by
removing different combinations of higher \Kstar and $\rho$ resonances 
(with low fit fractions), or changing the functional form of the resonance shapes, 
has little effect on the total $\chi^2$ of the fit, or on the values of the \CP parameters 
(at most $1^{\circ}$ for $\gamma$ and $0.002$ for $\kappa\rs$).  
As an extreme we consider a model without the $\sigma_1$ and/or $\sigma_2$ 
scalar resonances, or the CLEO Breit-Wigner model~\cite{ref:cleomodel}. Fits to these models result in a significantly 
larger $\chi^2$ than that of the nominal model, but the effect on the \CP parameters are still small,
as indicated in Table~\ref{tab:cartesian-syst}. These uncertainties translate to $\kappa\rs$, $\gamma$ and \deltas as
0.05, $9^\circ$, and $11^\circ$, 
respectively~\footnote{The Dalitz model uncertainties on $\kappa\rs$, $\gamma$ and \deltas are obtained by repeating the
fits to the same high statistics pseudo-experiments but now fitting directly to these \CP parameters. This method
accounts for the systematic correlations among the Cartesian \CP-violating parameters.}.
For $\Bm \to \dodstartilde \Km$ decays,
the same procedure to propagate the uncertainties on \rbbs, $\gamma$, and \deltabbs gives
$0.027(0.027)$, $11^\circ$, and $14^\circ(13^\circ)$, respectively.

As an additional cross-check of the fact that models without the $\sigma$ scalars are in fact an extreme case, we repeated
the above procedure using as alternative model the $\pi\pi$ S-wave K-matrix model described in Sec.~\ref{sec:dalitzModel} instead of
the CLEO Breit-Wigner model (the latter used to quote the Dalitz model systematic uncertainty, as described before). 
The effect on the Cartesian \CP parameters is more than three times smaller, as shown
in Table~\ref{tab:cartesian-syst}. These uncertainties translate to $\kappa\rs$, $\gamma$ and \deltas as
0.015, $3^\circ$, and $2^\circ$, respectively. For $\Bm \to \dodstartilde \Km$ decays, the scaling of the 
effect on the Cartesian \CP parameters is similar to $\Bm \to D \Kstarm$. In terms of \rbbs, $\gamma$, and \deltabbs
the variation is $0.004(0.003)$, $2^\circ$, and $2^\circ(2^\circ)$, respectively.

\section{INTERPRETATION AND RESULTS}
\label{sec:frequentist}

\begin{figure}[!t]
\begin{center}   
\begin{tabular} {c} 
 \includegraphics[height=8cm]{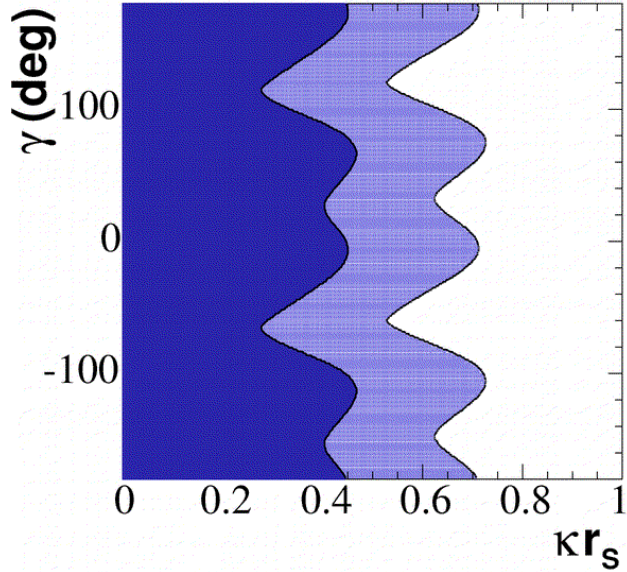}
\end{tabular}   
\caption{Two-dimensional projection onto the $\kappa\rs-\g$ plane of the three-dimensional one- (dark) and two- (light)
standard deviation regions, including statistical and experimental systematic uncertainties, for $\Bm \to \Dztilde \Kstarm$.
}
\label{fig:contours-rsvsgamma-dkstOnly}
\end{center} 
\end{figure}

A frequentist (Neyman) analysis~\cite{ref:pdg2004} has been adopted to interpret the constraints 
on $\zpm \equiv (\xspm,\yspm)$ in terms of $\p \equiv (\kappa\rs,\deltas,\g)$. We construct an analytical 
parameterization of the four-dimensional probability density function ${\cal P}$ of \zpm as a function of \p,
\bea
\frac{d^4 {\cal P}}{d^2 \zplus d^2 \zminus} (\zplus,\zminus | \p ) & = & \frac{d^2 {\cal P}}{d^2 \zplus} (\zplus | \p )
   \frac{d^2 {\cal P}}{d^2 \zminus} (\zminus | \p )~, 
\label{eq:d4pdf}
\eea
where
\bea
\frac{d^2 {\cal P}}{d^2 \zpm} (\zpm | \p ) & = & 
   G_2 \left[ \zpm;\kappa\rs\cos(\deltas\pm\gamma),\kappa\rs\sin(\deltas\pm\gamma),\sigma_{x_{s\pm}},\sigma_{y_{s\pm}},\rho_{s\pm} \right]
\label{eq:d2pdf}
\eea
is a two-dimensional Gaussian distribution~\cite{ref:2dGauss}. Here, $\rho_{s\pm}$ represents the correlation between $x_{s\pm}$ and $y_{s\pm}$.
For a given \p, the three-dimensional confidence level is estimated as ${\cal C}(\p) = 1 - \alpha(\p)$, where $\alpha(\p)$ 
is calculated by integrating Eq.~(\ref{eq:d4pdf}) over a domain defined by all points in 
the four-dimensional fit parameter space closer (larger PDF) to $\p$ than the fitted data values,
\bea
\frac{d^4 {\cal P}}{d^2 \zplus d^2 \zminus} (\zplus,\zminus | \p ) \ge \frac{d^4 {\cal P}}{d^2 \zplus d^2 \zminus} (\zpdata,\zmdata | \p )~.
\label{eq:Ddef}
\eea
The one (two) standard deviation region of the \CP parameters is defined as the set of \p values 
(constructed from a large number of pseudo-experiments) for which $\alpha(\p)$ is smaller than 19.88\% (73.85\%).
Figure~\ref{fig:contours-rsvsgamma-dkstOnly} shows the constraints in the $\kappa\rs-\g$ plane as obtained by projecting
the three-dimensional confidence regions, including statistical and experimental systematic uncertainties.

\begin{figure}[!t]
\begin{center}   
\begin{tabular} {ccc} 
 \includegraphics[height=4.9cm]{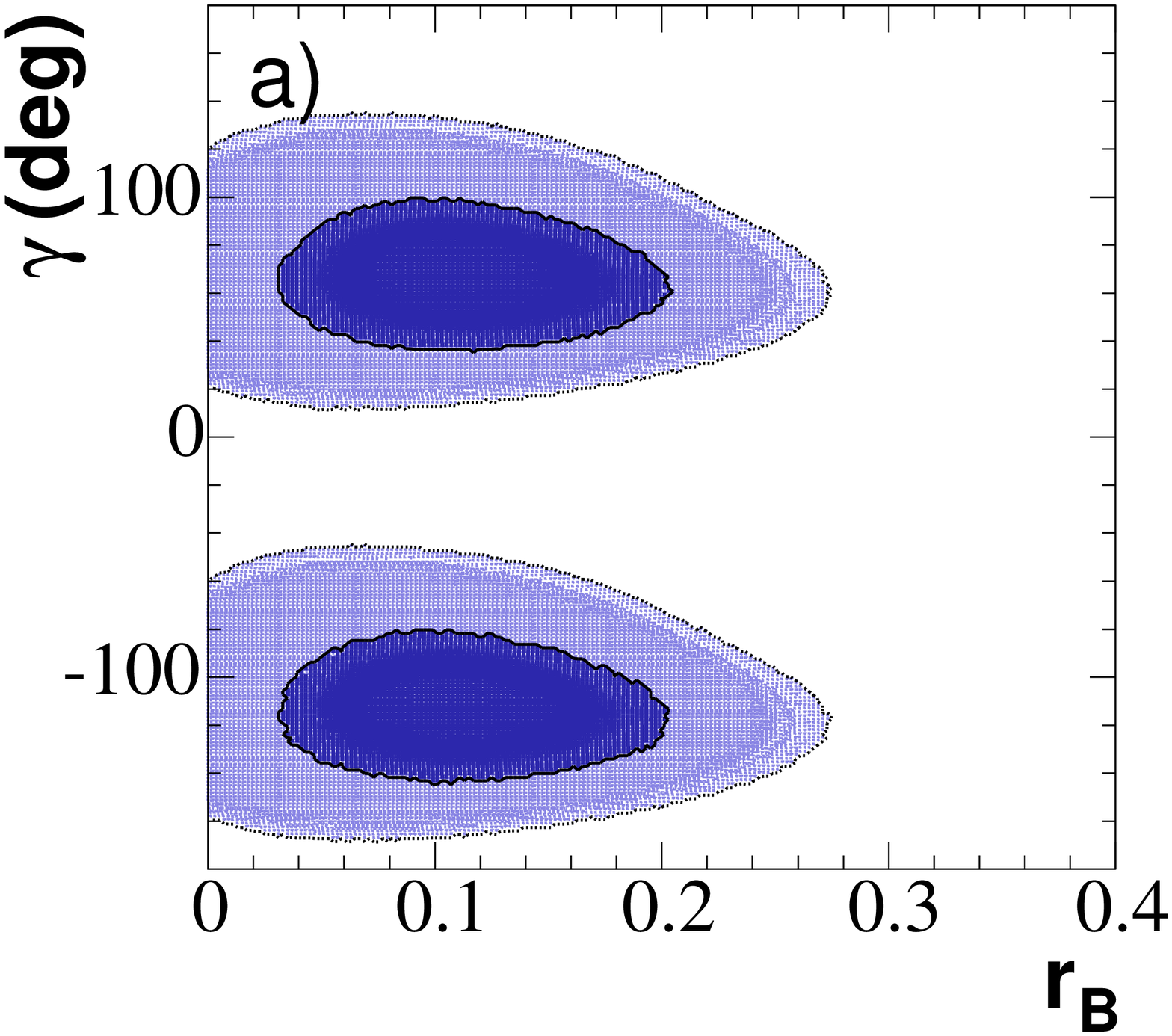} &
 \includegraphics[height=4.9cm]{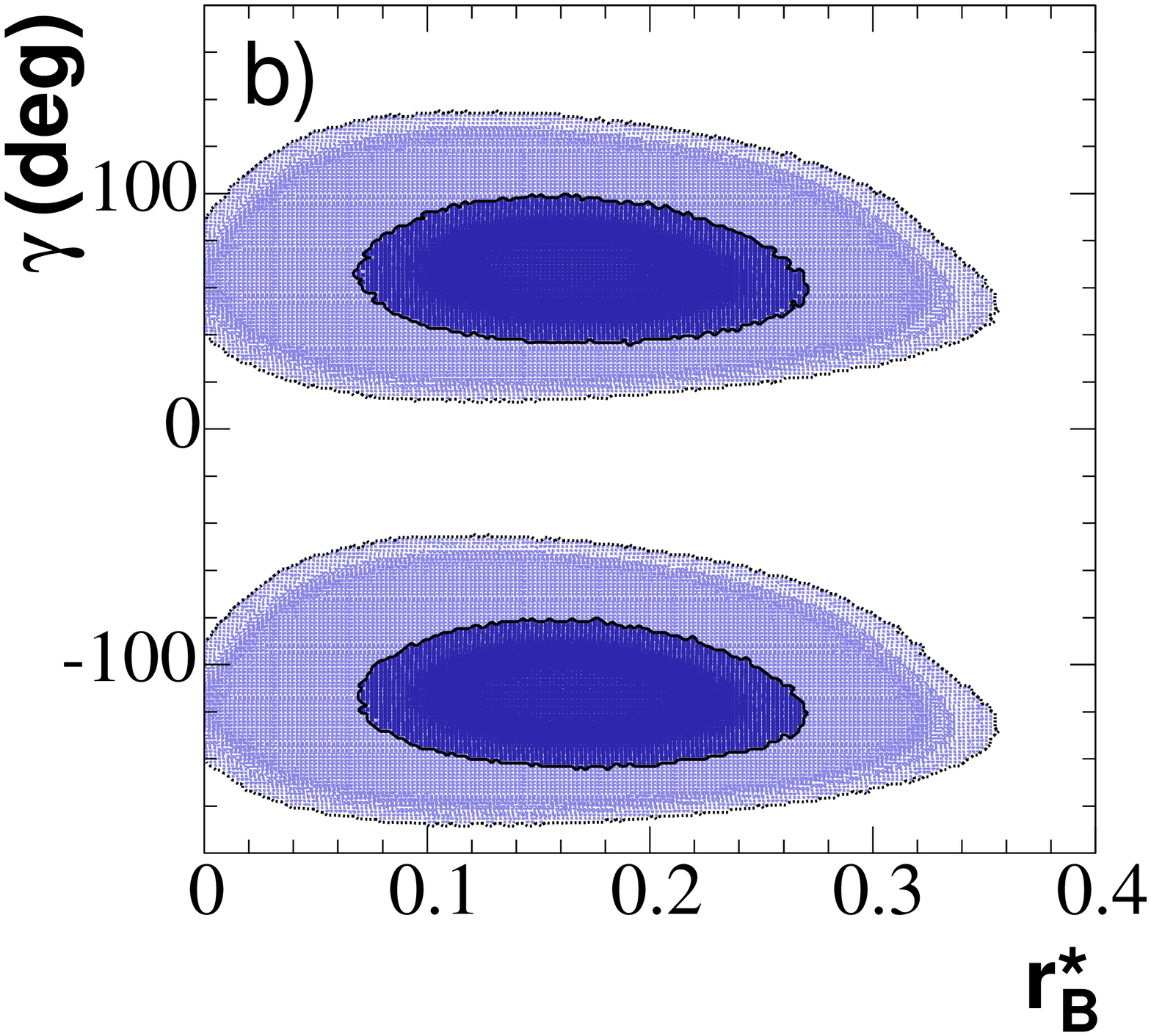} &
 \includegraphics[height=4.9cm]{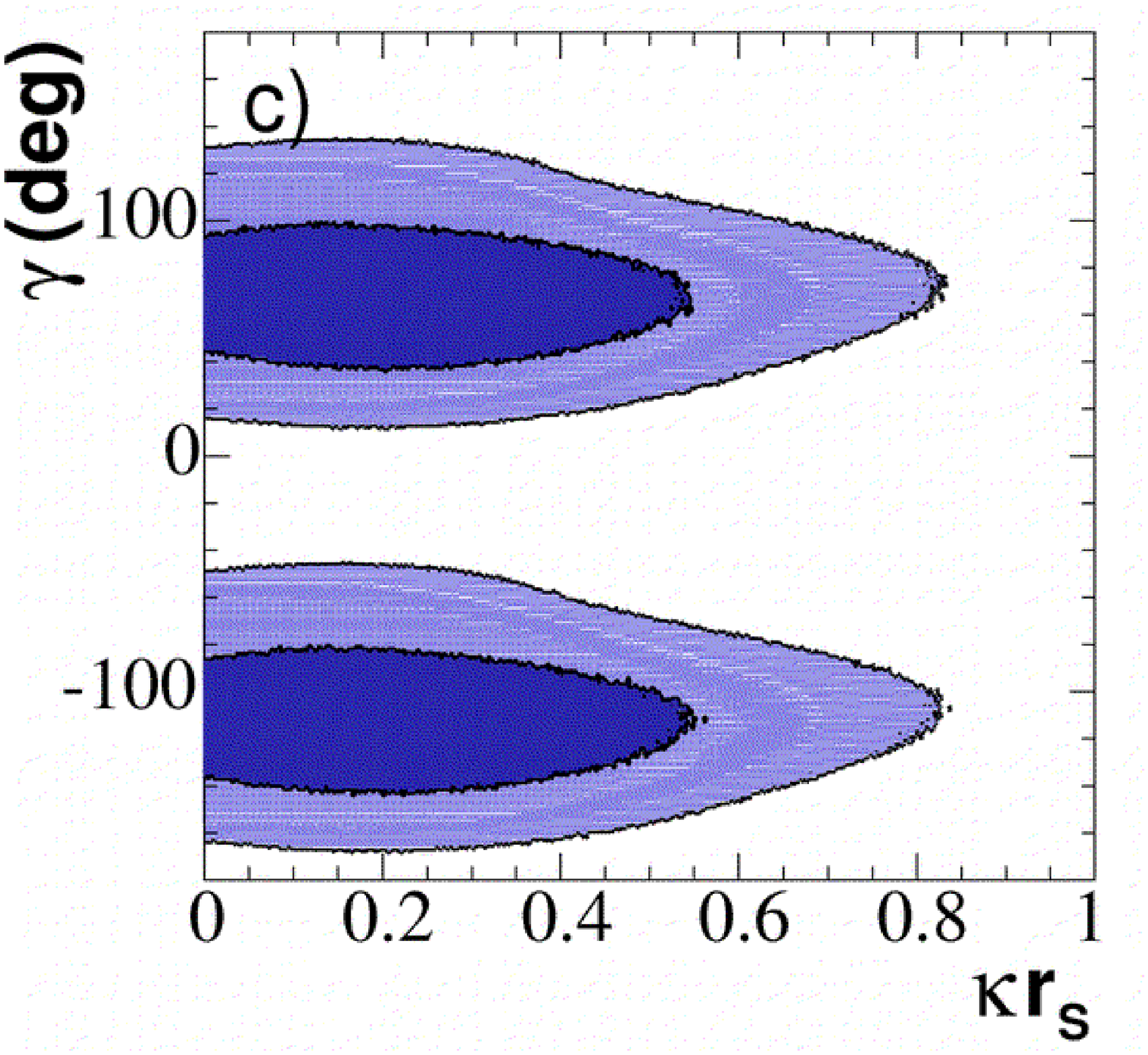} 
\end{tabular}   
\caption{Two-dimensional projections onto the (a) $\rb-\g$, (b) $\rbs-\g$, 
and (c) $\kappa\rs-\g$ planes of the 
seven-dimensional one- (dark) and two- (light) standard deviation regions, for the combination
of $\Bm \to \dodstartilde \Km$ and $\Bm \to \Dztilde \Kstarm$ modes.
}
\label{fig:contours-rbvsgamma}
\end{center} 
\end{figure}

$\Bm \to \Dztilde \Kstarm$ events can be used in combination with $\Bm \to \dodstartilde \Km$~\cite{ref:babar_dalitz} to 
improve the overall constraints on $\gamma$. The procedure used to combine the three \B decay 
channels is identical to the one described above, but with increased number of dimensions.
In this case, the dimension of the fit parameter space is twelve, 
${\bf z_{\pm}} \equiv (\xbbspm,\ybbspm,\xspm,\yspm)$. The \CP parameter space has instead seven dimensions, 
$\p \equiv (\rbbs,\deltabbs,\kappa\rs,\deltas,\g)$.
The one (two) standard deviation region of the \CP parameters is defined as the set of $\p$ values 
for which $\alpha(\p)$ is larger than 0.52\% (22.02\%). Figure~\ref{fig:contours-rbvsgamma} shows the two-dimensional 
projections onto the $\rb-\g$, $\rbs-\g$, and $\kappa\rs-\g$ planes, including statistical and experimental systematic uncertainties.
The figures show that this Dalitz analysis has a two-fold ambiguity.
The combination if the three signal modes yields 
$\gamma = \reslinepolvalppp{67}{28}{13}{11}$, 
where the first error is statistical, 
the second is the experimental systematic uncertainty and the third reflects the Dalitz model uncertainty. 
Of the two possible solutions we choose the one with $0<\gamma<180^\circ$. 
The contribution to the Dalitz model uncertainty due to the description of the $\pi\pi$ S-wave in $\Dz \to \KS \pim \pip$
is $3^\circ$.
From this combination, $\kappa\rs$ is constrained
to be $<0.50~(0.75)$ at one (two) standard deviation level. It is worth noting that the value of $\kappa \rs$ depends on
the selected phase space region of $\Bm \to \Dztilde (\KS \pim)$ events without introducing any bias on the 
\mbox{extraction of $\gamma$.}
 
The constraint on \g is consistent with that reported by the Belle Collaboration~\cite{ref:belle_dalitz,ref:belle_btodkstar}.
However, the statistical error turns out to be
larger than that of Belle because our data favors smaller values of
\rbbs and $\kappa\rs$. Simulation studies confirm that the
difference in statistical errors is consistent with the scaling
expected from the different actual values.

\section{ACKNOWLEDGMENTS}
\label{sec:Acknowledgments}
We are grateful for the 
extraordinary contributions of our \pep2\ colleagues in
achieving the excellent luminosity and machine conditions
that have made this work possible.
The success of this project also relies critically on the 
expertise and dedication of the computing organizations that 
support \babar.
The collaborating institutions wish to thank 
SLAC for its support and the kind hospitality extended to them. 
This work is supported by the
US Department of Energy
and National Science Foundation, the
Natural Sciences and Engineering Research Council (Canada),
Institute of High Energy Physics (China), the
Commissariat \`a l'Energie Atomique and
Institut National de Physique Nucl\'eaire et de Physique des Particules
(France), the
Bundesministerium f\"ur Bildung und Forschung and
Deutsche Forschungsgemeinschaft
(Germany), the
Istituto Nazionale di Fisica Nucleare (Italy),
the Foundation for Fundamental Research on Matter (The Netherlands),
the Research Council of Norway, the
Ministry of Science and Technology of the Russian Federation, and the
Particle Physics and Astronomy Research Council (United Kingdom). 
Individuals have received support from 
CONACyT (Mexico),
the A. P. Sloan Foundation, 
the Research Corporation,
and the Alexander von Humboldt Foundation.


\begin{thebibliography}{99}
\bibitem{ref:FOCUS} FOCUS Collaboration, J.~M.~Link {\em et al.}, \plb{585}, 200 (2004). 
One should note that there is a misprint of the $(1-s_{A0})$ factor in the Adler zero expression.
\end{thebibliography}
\end{document}